\documentclass[10pt]{article}
\usepackage[left=1in,top=1in,right=1in,bottom=1in]{geometry} 

\usepackage{authblk} 
\usepackage{xurl} 
\usepackage{graphicx}
\usepackage{amsfonts} 
\usepackage{amssymb}
\usepackage{amsmath} 
\usepackage{graphicx}
\usepackage{booktabs} 
\usepackage{multirow} 
\usepackage{tablefootnote} 
\usepackage{caption} 
\usepackage{subcaption}  
\usepackage{acronym} 
\usepackage[table,xcdraw]{xcolor} 
\usepackage{hyperref} 
\usepackage{fancyvrb} 
\usepackage[square,numbers]{natbib} 
\usepackage{tablefootnote} 

\usepackage{blindtext}
\usepackage{diagbox} 

\definecolor{td}{HTML}{f44336}
\definecolor{ia}{HTML}{9c27b0}
\definecolor{dbw}{HTML}{3f51b5}
\definecolor{lcl}{HTML}{03a9f4}
\definecolor{cl}{HTML}{009688}
\definecolor{mm}{HTML}{e91e63}
\definecolor{jgs}{HTML}{ff9800}

\newcommand*\samethanks[1][\value{footnote}]{\footnotemark[#1]}

\begin{document}

\title{University of Copenhagen Participation in TREC Health Misinformation Track 2020}

\author[]{Lucas Chaves Lima\thanks{The first and second author contributed equally.}}
\author[]{Dustin Brandon Wright\samethanks}
\author[]{Isabelle Augenstein}
\author[]{Maria Maistro}

\affil[]{University of Copenhagen \\ \textit {\{lcl,dw,augenstein,mm\}@di.ku.dk}}

\date{}

\maketitle

\acrodef{3G}[3G]{Third Generation Mobile System}
\acrodef{5S}[5S]{Streams, Structures, Spaces, Scenarios, Societies}
\acrodef{AAAI}[AAAI]{Association for the Advancement of Artificial Intelligence}
\acrodef{AAL}[AAL]{Annotation Abstraction Layer}
\acrodef{AAM}[AAM]{Automatic Annotation Manager}
\acrodef{ACLIA}[ACLIA]{Advanced Cross-Lingual Information Access}
\acrodef{ACM}[ACM]{Association for Computing Machinery}
\acrodef{ADSL}[ADSL]{Asymmetric Digital Subscriber Line}
\acrodef{ADUI}[ADUI]{ADministrator User Interface}
\acrodef{AIP}[AIP]{Archival Information Package}
\acrodef{AJAX}[AJAX]{Asynchronous JavaScript Technology and \acs{XML}}
\acrodef{ALU}[ALU]{Aritmetic-Logic Unit}
\acrodef{AMUSID}[AMUSID]{Adaptive MUSeological IDentity-service}
\acrodef{ANOVA}[ANOVA]{ANalysis Of VAriance}
\acrodef{ANSI}[ANSI]{American National Standards Institute}
\acrodef{AP}[AP]{Average Precision}
\acrodef{APC}[APC]{AP Correlation}
\acrodef{API}[API]{Application Program Interface}
\acrodef{AR}[AR]{Address Register}
\acrodef{AS}[AS]{Annotation Service}
\acrodef{ASAP}[ASAP]{Adaptable Software Architecture Performance}
\acrodef{ASI}[ASI]{Annotation Service Integrator}
\acrodef{ASL}[ASL]{Achieved Significance Level}
\acrodef{ASM}[ASM]{Annotation Storing Manager}
\acrodef{ASR}[ASR]{Automatic Speech Recognition}
\acrodef{ASUI}[ASUI]{ASsessor User Interface}
\acrodef{ATIM}[ATIM]{Annotation Textual Indexing Manager}
\acrodef{AUC}[AUC]{Area Under the ROC Curve}
\acrodef{AUI}[AUI]{Administrative User Interface}
\acrodef{AWARE}[AWARE]{Assessor-driven Weighted Averages for Retrieval Evaluation}
\acrodef{BANKS-I}[BANKS-I]{Browsing ANd Keyword Searching I}
\acrodef{BANKS-II}[BANKS-II]{Browsing ANd Keyword Searching II}
\acrodef{bpref}[bpref]{Binary Preference}
\acrodef{BNF}[BNF]{Backus and Naur Form}
\acrodef{BRICKS}[BRICKS]{Building Resources for Integrated Cultural Knowledge Services}
\acrodef{CAM}[CAM]{Convex Aggregating Measure}
\acrodef{CAN}[CAN]{Content Addressable Netword}
\acrodef{CAS}[CAS]{Content-And-Structure}
\acrodef{CBSD}[CBSD]{Component-Based Software Developlement}
\acrodef{CBSE}[CBSE]{Component-Based Software Engineering}
\acrodef{CB-SPE}[CB-SPE]{Component-Based \acs{SPE}}
\acrodef{CD}[CD]{Collaboration Diagram}
\acrodef{CD}[CD]{Compact Disk}
\acrodef{CENL}[CENL]{Conference of European National Librarians}
\acrodef{CIDOC CRM}[CIDOC CRM]{CIDOC Conceptual Reference Model}
\acrodef{CIR}[CIR]{Current Instruction Register}
\acrodef{CIRCO}[CIRCO]{Coordinated Information Retrieval Components Orchestration}
\acrodef{CG}[CG]{Cumulated Gain}
\acrodef{CL}[CL]{Curriculum Learning}
\acrodef{CLEF1}[CLEF]{Cross-Language Evaluation Forum}
\acrodef{CLEF}[CLEF]{Conference and Labs of the Evaluation Forum}
\acrodef{CLIR}[CLIR]{Cross Language Information Retrieval}
\acrodef{CM}[CM]{Continuation Methods}
\acrodef{CMS}[CMS]{Content Management System}
\acrodef{CMT}[CMT]{Campaign Management Tool}
\acrodef{CNR}[CNR]{Italian National Council of Research}
\acrodef{CO}[CO]{Content-Only}
\acrodef{COD}[COD]{Code On Demand}
\acrodef{CODATA}[CODATA]{Committee on Data for Science and Technology}
\acrodef{COLLATE}[COLLATE]{Collaboratory for Annotation Indexing and Retrieval of Digitized Historical Archive Material}
\acrodef{CP}[CP]{Characteristic Pattern}
\acrodef{CPE}[CPE]{Control Processor Element}
\acrodef{CPU}[CPU]{Central Processing Unit}
\acrodef{CQL}[CQL]{Contextual Query Language}
\acrodef{CRP}[CRP]{Cumulated Relative Position}
\acrodef{CRUD}[CRUD]{Create--Read--Update--Delete}
\acrodef{CS}[CS]{Characteristic Structure}
\acrodef{CSM}[CSM]{Campaign Storing Manager}
\acrodef{CSS}[CSS]{Cascading Style Sheets}
\acrodef{CU}[CU]{Control Unit}
\acrodef{CUI}[CUI]{Client User Interface}
\acrodef{CV}[CV]{Cross-Validation}
\acrodef{DAFFODIL}[DAFFODIL]{Distributed Agents for User-Friendly Access of Digital Libraries}
\acrodef{DAO}[DAO]{Data Access Object}
\acrodef{DARE}[DARE]{Drawing Adequate REpresentations}
\acrodef{DARPA}[DARPA]{Defense Advanced Research Projects Agency}
\acrodef{DAS}[DAS]{Distributed Annotation System}
\acrodef{DB}[DB]{DataBase}
\acrodef{DBMS}[DBMS]{DataBase Management System}
\acrodef{DC}[DC]{Dublin Core}
\acrodef{DCG}[DCG]{Discounted Cumulated Gain}
\acrodef{DCMI}[DCMI]{Dublin Core Metadata Initiative}
\acrodef{DCV}[DCV]{Document Cut--off Value}
\acrodef{DD}[DD]{Deployment Diagram}
\acrodef{DDC}[DDC]{Dewey Decimal Classification}
\acrodef{DDS}[DDS]{Direct Data Structure}
\acrodef{DF}[DF]{Degrees of Freedom}
\acrodef{DFI}[DFI]{Divergence From Independence}
\acrodef{DFR}[DFR]{Divergence From Randomness}
\acrodef{DHT}[DHT]{Distributed Hash Table}
\acrodef{DI}[DI]{Digital Image}
\acrodef{DIKW}[DIKW]{Data, Information, Knowledge, Wisdom}
\acrodef{DIL}[DIL]{\acs{DIRECT} Integration Layer}
\acrodef{DiLAS}[DiLAS]{Digital Library Annotation Service}
\acrodef{DIRECT}[DIRECT]{Distributed Information Retrieval Evaluation Campaign Tool}
\acrodef{DKMS}[DKMS]{Data and Knowledge Management System}
\acrodef{DL}[DL]{Digital Library}
\acrodefplural{DL}[DL]{Digital Libraries}
\acrodef{DLMS}[DLMS]{Digital Library Management System}
\acrodef{DLOG}[DL]{Description Logics}
\acrodef{DLS}[DLS]{Digital Library System}
\acrodef{DLSS}[DLSS]{Digital Library Service System}
\acrodef{DM}[DM]{Data Mining}
\acrodef{DO}[DO]{Digital Object}
\acrodef{DOI}[DOI]{Digital Object Identifier}
\acrodef{DOM}[DOM]{Document Object Model}
\acrodef{DoMDL}[DoMDL]{Document Model for Digital Libraries}
\acrodef{DP}[DP]{Discriminative Power}
\acrodef{DPBF}[DPBF]{Dynamic Programming Best-First}
\acrodef{DR}[DR]{Data Register}
\acrodef{DRIVER}[DRIVER]{Digital Repository Infrastructure Vision for European Research}
\acrodef{DTD}[DTD]{Document Type Definition}
\acrodef{DVD}[DVD]{Digital Versatile Disk}
\acrodef{EAC-CPF}[EAC-CPF]{Encoded Archival Context for Corporate Bodies, Persons, and Families}
\acrodef{EAD}[EAD]{Encoded Archival Description}
\acrodef{EAN}[EAN]{International Article Number}
\acrodef{E-A-T}[E-A-T]{Expertise, Authoritativeness, and Trustworthiness}
\acrodef{ECD}[ECD]{Enhanced Contenty Delivery}
\acrodef{ECDL}[ECDL]{European Conference on Research and Advanced Technology for Digital Libraries}
\acrodef{EDM}[EDM]{Europeana Data Model}
\acrodef{EG}[EG]{Execution Graph}
\acrodef{ELDA}[ELDA]{Evaluation and Language resources Distribution Agency}
\acrodef{ELRA}[ELRA]{European Language Resources Association}
\acrodef{EM}[EM]{Expectation Maximization}
\acrodef{EMMA}[EMMA]{Extensible MultiModal Annotation}
\acrodef{EPROM}[EPROM]{Erasable Programmable \acs{ROM}}
\acrodef{EQNM}[EQNM]{Extended Queueing Network Model}
\acrodef{ER}[ER]{Entity--Relationship}
\acrodef{ERR}[ERR]{Expected Reciprocal Rank}
\acrodef{ETL}[ETL]{Extract-Transform-Load}
\acrodef{FAST}[FAST]{Flexible Annotation Service Tool}
\acrodef{FIFO}[FIFO]{First-In / First-Out}
\acrodef{FIRE}[FIRE]{Forum for Information Retrieval Evaluation}
\acrodef{FN}[FN]{False Negative}
\acrodef{FNR}[FNR]{False Negative Rate}
\acrodef{FOAF}[FOAF]{Friend of a Friend}
\acrodef{FORESEE}[FORESEE]{FOod REcommentation sErvER}
\acrodef{FP}[FP]{False Positive}
\acrodef{FPR}[FPR]{False Positive Rate}
\acrodef{GIF}[GIF]{Graphics Interchange Format}
\acrodef{GIR}[GIR]{Geografic Information Retrieval}
\acrodef{GAP}[GAP]{Graded Average Precision}
\acrodef{GLM}[GLM]{General Linear Model}
\acrodef{GLMM}[GLMM]{General Linear Mixed Model}
\acrodef{GMAP}[GMAP]{Geometric Mean Average Precision}
\acrodef{GoP}[GoP]{Grid of Points}
\acrodef{GPRS}[GPRS]{General Packet Radio Service}
\acrodef{gP}[gP]{Generalized Precision}
\acrodef{gR}[gR]{Generalized Recall}
\acrodef{gRBP}[gRBP]{Graded Rank-Biased Precision}
\acrodef{GTIN}[GTIN]{Global Trade Item Number}
\acrodef{GUI}[GUI]{Graphical User Interface}
\acrodef{GW}[GW]{Gateway}
\acrodef{HCI}[HCI]{Human Computer Interaction}
\acrodef{HDS}[HDS]{Hybrid Data Structure}
\acrodef{HIR}[HIR]{Hypertext Information Retrieval}
\acrodef{HIT}[HIT]{Human Intelligent Task}
\acrodef{HITS}[HITS]{Hyperlink-Induced Topic Search}
\acrodef{HTML}[HTML]{HyperText Markup Language}
\acrodef{HTTP}[HTTP]{HyperText Transfer Protocol}
\acrodef{HSD}[HSD]{Honestly Significant Difference}
\acrodef{ICA}[ICA]{International Council on Archives}
\acrodef{ICSU}[ICSU]{International Council for Science}
\acrodef{IDF}[IDF]{Inverse Document Frequency}
\acrodef{IDS}[IDS]{Inverse Data Structure}
\acrodef{IEEE}[IEEE]{Institute of Electrical and Electronics Engineers}
\acrodef{IEI}[IEI]{Istituto della Enciclopedia Italiana fondata da Giovanni Treccani}
\acrodef{IETF}[IETF]{Internet Engineering Task Force}
\acrodef{IMS}[IMS]{Information Management System}
\acrodef{IMSPD}[IMS]{Information Management Systems Research Group}
\acrodef{indAP}[indAP]{Induced Average Precision}
\acrodef{infAP}[infAP]{Inferred Average Precision}
\acrodef{INEX}[INEX]{INitiative for the Evaluation of \acs{XML} Retrieval}
\acrodef{INS-M}[INS-M]{Inverse Set Data Model}
\acrodef{INTR}[INTR]{Interrupt Register}
\acrodef{IP}[IP]{Internet Protocol}
\acrodef{IPSA}[IPSA]{Imaginum Patavinae Scientiae Archivum}
\acrodef{IR}[IR]{Information Retrieval}
\acrodef{IRON}[IRON]{Information Retrieval ON}
\acrodef{IRON2}[IRON$^2$]{Information Retrieval On aNNotations}
\acrodef{IRON-SAT}[IRON-SAT]{\acs{IRON} - Statistical Analysis Tool}
\acrodef{IRS}[IRS]{Information Retrieval System}
\acrodef{ISAD(G)}[ISAD(G)]{International Standard for Archival Description (General)}
\acrodef{ISBN}[ISBN]{International Standard Book Number}
\acrodef{ISIS}[ISIS]{Interactive SImilarity Search}
\acrodef{ISJ}[ISJ]{Interactive Searching and Judging}
\acrodef{ISO}[ISO]{International Organization for Standardization}
\acrodef{ITU}[ITU]{International Telecommunication Union }
\acrodef{ITU-T}[ITU-T]{Telecommunication Standardization Sector of \acs{ITU}}
\acrodef{IV}[IV]{Information Visualization}
\acrodef{JAN}[JAN]{Japanese Article Number}
\acrodef{JDBC}[JDBC]{Java DataBase Connectivity}
\acrodef{JMB}[JMB]{Java--Matlab Bridge}
\acrodef{JPEG}[JPEG]{Joint Photographic Experts Group}
\acrodef{JSON}[JSON]{JavaScript Object Notation}
\acrodef{JSP}[JSP]{Java Server Pages}
\acrodef{JTE}[JTE]{Java-Treceval Engine}
\acrodef{KDE}[KDE]{Kernel Density Estimation}
\acrodef{KLD}[KLD]{Kullback-Leibler Divergence}
\acrodef{KLAPER}[KLAPER]{Kernel LAnguage for PErformance and Reliability analysis}
\acrodef{LAM}[LAM]{Libraries, Archives, and Museums}
\acrodef{LAM2}[LAM]{Logistic Average Misclassification}
\acrodef{LAN}[LAN]{Local Area Network}
\acrodef{LD}[LD]{Linked Data}
\acrodef{LEAF}[LEAF]{Linking and Exploring Authority Files}
\acrodef{LIDO}[LIDO]{Lightweight Information Describing Objects}
\acrodef{LIFO}[LIFO]{Last-In / First-Out}
\acrodef{LM}[LM]{Language Model}
\acrodef{LMT}[LMT]{Log Management Tool}
\acrodef{LOD}[LOD]{Linked Open Data}
\acrodef{LODE}[LODE]{Linking Open Descriptions of Events}
\acrodef{LpO}[LpO]{Leave-$p$-Out}
\acrodef{LRE}[LRE]{Local Rank Error}
\acrodef{LRM}[LRM]{Local Relational Model}
\acrodef{LRU}[LRU]{Last Recently Used}
\acrodef{LS}[LS]{Lexical Signature}
\acrodef{LSM}[LSM]{Log Storing Manager}
\acrodef{LtR}[LtR]{Learning to Rank}
\acrodef{LUG}[LUG]{Lexical Unit Generator}
\acrodef{MA}[MA]{Mobile Agent}
\acrodef{MA}[MA]{Moving Average}
\acrodef{MACS}[MACS]{Multilingual ACcess to Subjects}
\acrodef{MADCOW}[MADCOW]{Multimedia Annotation of Digital Content Over the Web}
\acrodef{MAD}[MAD]{Mean Assessed Documents}
\acrodef{MADP}[MADP]{Mean Assessed Documents Precision}
\acrodef{MADS}[MADS]{Metadata Authority Description Standard}
\acrodef{MAP}[MAP]{Mean Average Precision}
\acrodef{MARC}[MARC]{Machine Readable Cataloging}
\acrodef{MATTERS}[MATTERS]{MATlab Toolkit for Evaluation of information Retrieval Systems}
\acrodef{MDA}[MDA]{Model Driven Architecture}
\acrodef{MDD}[MDD]{Model-Driven Development}
\acrodef{METS}[METS]{Metadata Encoding and Transmission Standard}
\acrodef{MIDI}[MIDI]{Musical Instrument Digital Interface}
\acrodef{MIME}[MIME]{Multipurpose Internet Mail Extensions}
\acrodef{ML}[ML]{Machine Learning}
\acrodef{MLIA}[MLIA]{MultiLingual Information Access}
\acrodef{MM}[MM]{Machinery Model}
\acrodef{MMU}[MMU]{Memory Management Unit}
\acrodef{MODS}[MODS]{Metadata Object Description Schema}
\acrodef{MOF}[MOF]{Meta-Object Facility}
\acrodef{MP}[MP]{Markov Precision}
\acrodef{MPEG}[MPEG]{Motion Picture Experts Group}
\acrodef{MRD}[MRD]{Machine Readable Dictionary}
\acrodef{MRF}[MRF]{Markov Random Field}
\acrodef{MS}[MS]{Mean Squares}
\acrodef{MSAC}[MSAC]{Multilingual Subject Access to Catalogues}
\acrodef{MSE}[MSE]{Mean Square Error}
\acrodef{MT}[MT]{Machine Translation}
\acrodef{MV}[MV]{Majority Vote}
\acrodef{MVC}[MVC]{Model-View-Controller}
\acrodef{NACSIS}[NACSIS]{NAtional Center for Science Information Systems}
\acrodef{NAP}[NAP]{Network processors Applications Profile}
\acrodef{NCP}[NCP]{Normalized Cumulative Precision}
\acrodef{nCG}[nCG]{Normalized Cumulated Gain}
\acrodef{nCRP}[nCRP]{Normalized Cumulated Relative Position}
\acrodef{nDCG}[nDCG]{Normalized Discounted Cumulated Gain}
\acrodef{nDE}[nDE]{Normalized Discounted Error}
\acrodef{NESTOR}[NESTOR]{NEsted SeTs for Object hieRarchies}
\acrodef{NEXI}[NEXI]{Narrowed Extended XPath I}
\acrodef{NII}[NII]{National Institute of Informatics}
\acrodef{NISO}[NISO]{National Information Standards Organization}
\acrodef{NIST}[NIST]{National Institute of Standards and Technology}
\acrodef{NLP}[NLP]{Natural Language Processing}
\acrodef{NLRE}[NLRE]{Normalized Local Rank Error}
\acrodef{NN}[NN]{Neural Network}
\acrodef{NP}[NP]{Network Processor}
\acrodef{NR}[NR]{Normalized Recall}
\acrodef{NS-M}[NS-M]{Nested Set Model}
\acrodef{NTCIR}[NTCIR]{NII Testbeds and Community for Information access Research}
\acrodef{OAI}[OAI]{Open Archives Initiative}
\acrodef{OAI-ORE}[OAI-ORE]{Open Archives Initiative Object Reuse and Exchange}
\acrodef{OAI-PMH}[OAI-PMH]{Open Archives Initiative Protocol for Metadata Harvesting}
\acrodef{OAIS}[OAIS]{Open Archival Information System}
\acrodef{OC}[OC]{Operation Code}
\acrodef{OCLC}[OCLC]{Online Computer Library Center}
\acrodef{OMG}[OMG]{Object Management Group}
\acrodef{OO}[OO]{Object Oriented}
\acrodef{OODB}[OODB]{Object-Oriented \acs{DB}}
\acrodef{OODBMS}[OODBMS]{Object-Oriented \acs{DBMS}}
\acrodef{OPAC}[OPAC]{Online Public Access Catalog}
\acrodef{OQL}[OQL]{Object Query Language}
\acrodef{ORP}[ORP]{Open Relevance Project}
\acrodef{OSIRIS}[OSIRIS]{Open Service Infrastructure for Reliable and Integrated process Support}
\acrodef{P}[P]{Precision}
\acrodef{P2P}[P2P]{Peer-To-Peer}
\acrodef{PA}[PA]{Performance Analysis}
\acrodef{PAMT}[PAMT]{Pool-Assessment Management Tool}
\acrodef{PASM}[PASM]{Pool-Assessment Storing Manager}
\acrodef{PC}[PC]{Program Counter}
\acrodef{PCP}[PCP]{Pre-Commercial Procurement}
\acrodef{PCR}[PCR]{Peripherical Command Register}
\acrodef{PDA}[PDA]{Personal Digital Assistant}
\acrodef{PDF}[PDF]{Probability Density Function}
\acrodef{PDR}[PDR]{Peripherical Data Register}
\acrodef{PIR}[PIR]{Personalized Information Retrieval}
\acrodef{POI}[POI]{\acs{PURL}-based Object Identifier}
\acrodef{PoS}[PoS]{Part of Speech}
\acrodef{PPE}[PPE]{Programmable Processing Engine}
\acrodef{PREFORMA}[PREFORMA]{PREservation FORMAts for culture information/e-archives}
\acrodef{PRIMAmob-UML}[PRIMAmob-UML]{mobile \acs{PRIMA-UML}}
\acrodef{PRIMA-UML}[PRIMA-UML]{PeRformance IncreMental vAlidation in \acs{UML}}
\acrodef{PROM}[PROM]{Programmable \acs{ROM}}
\acrodef{PROMISE}[PROMISE]{Participative Research labOratory  for Multimedia and Multilingual Information Systems Evaluation}
\acrodef{pSQL}[pSQL]{propagate \acs{SQL}}
\acrodef{PUI}[PUI]{Participant User Interface}
\acrodef{PURL}[PURL]{Persistent \acs{URL}}
\acrodef{QA}[QA]{Question Answering}
\acrodef{QoS-UML}[QoS-UML]{\acs{UML} Profile for QoS and Fault Tolerance}
\acrodef{QPA}[QPA]{Query Performance Analyzer}
\acrodef{R}[R]{Recall}
\acrodef{RAM}[RAM]{Random Access Memory}
\acrodef{RAMM}[RAM]{Random Access Machine}
\acrodef{RBO}[RBO]{Rank-Biased Overlap}
\acrodef{RBP}[RBP]{Rank-Biased Precision}
\acrodef{RDBMS}[RDBMS]{Relational \acs{DBMS}}
\acrodef{RDF}[RDF]{Resource Description Framework}
\acrodef{REST}[REST]{REpresentational State Transfer}
\acrodef{REV}[REV]{Remote Evaluation}
\acrodef{RFC}[RFC]{Request for Comments}
\acrodef{RIA}[RIA]{Reliable Information Access}
\acrodef{RMSE}[RMSE]{Root Mean Square Error}
\acrodef{RMT}[RMT]{Run Management Tool}
\acrodef{ROM}[ROM]{Read Only Memory}
\acrodef{ROMIP}[ROMIP]{Russian Information Retrieval Evaluation Seminar}
\acrodef{RoMP}[RoMP]{Rankings of Measure Pairs}
\acrodef{RoS}[RoS]{Rankings of Systems}
\acrodef{RP}[RP]{Relative Position}
\acrodef{RR}[RR]{Reciprocal Rank}
\acrodef{RSM}[RSM]{Run Storing Manager}
\acrodef{RST}[RST]{Rhetorical Structure Theory}
\acrodef{RT-UML}[RT-UML]{\acs{UML} Profile for Schedulability, Performance and Time}
\acrodef{SA}[SA]{Software Architecture}
\acrodef{SAL}[SAL]{Storing Abstraction Layer}
\acrodef{SAMT}[SAMT]{Statistical Analysis Management Tool}
\acrodef{SAN}[SAN]{Sistema Archivistico Nazionale}
\acrodef{SASM}[SASM]{Statistical Analysis Storing Manager}
\acrodef{SD}[SD]{Sequence Diagram}
\acrodef{SE}[SE]{Search Engine}
\acrodef{SEBD}[SEBD]{Convegno Nazionale su Sistemi Evoluti per Basi di Dati}
\acrodef{SERP}[SERP]{Search Engine Result Page}
\acrodef{SFT}[SFT]{Satisfaction--Frustration--Total}
\acrodef{SIL}[SIL]{Service Integration Layer}
\acrodef{SIP}[SIP]{Submission Information Package}
\acrodef{SKOS}[SKOS]{Simple Knowledge Organization System}
\acrodef{SM}[SM]{Software Model}
\acrodef{SME}[SME]{Statistics--Metrics-Experiments}
\acrodef{SMART}[SMART]{System for the Mechanical Analysis and Retrieval of Text}
\acrodef{SoA}[SoA]{Service-oriented Architectures}
\acrodef{SOA}[SOA]{Strength of Association}
\acrodef{SOAP}[SOAP]{Simple Object Access Protocol}
\acrodef{SOM}[SOM]{Self-Organizing Map}
\acrodef{SPARQL}[SPARQL]{Simple Protocol and RDF Query Language}
\acrodef{SPE}[SPE]{Software Performance Engineering}
\acrodef{SPINA}[SPINA]{Superimposed Peer Infrastructure for iNformation Access}
\acrodef{SPLIT}[SPLIT]{Stemming Program for Language Independent Tasks}
\acrodef{SPOOL}[SPOOL]{Simultaneous Peripheral Operations On Line}
\acrodef{SQL}[SQL]{Structured Query Language}
\acrodef{SR}[SR]{Sliding Ratio}
\acrodef{SR}[SR]{Status Register}
\acrodef{SRU}[SRU]{Search/Retrieve via \acs{URL}}
\acrodef{SS}[SS]{Sum of Squares}
\acrodef{SSTF}[SSTF]{Shortest Seek Time First}
\acrodef{STAR}[STAR]{Steiner-Tree Approximation in Relationship graphs}
\acrodef{STON}[STON]{STemming ON}
\acrodef{SVM}[SVM]{Support Vector Machine}
\acrodef{TAC}[TAC]{Text Analysis Conference}
\acrodef{TBG}[TBG]{Time-Biased Gain}
\acrodef{TCP}[TCP]{Transmission Control Protocol}
\acrodef{TEL}[TEL]{The European Library}
\acrodef{TERRIER}[TERRIER]{TERabyte RetrIEveR}
\acrodef{TF}[TF]{Term Frequency}
\acrodef{TFR}[TFR]{True False Rate}
\acrodef{TLD}[TLD]{Top Level Domain}
\acrodef{TME}[TME]{Topics--Metrics-Experiments}
\acrodef{TN}[TN]{True Negative}
\acrodef{TO}[TO]{Transfer Object}
\acrodef{TP}[TP]{True Positve}
\acrodef{TPR}[TPR]{True Positive Rate}
\acrodef{TRAT}[TRAT]{Text Relevance Assessing Task}
\acrodef{TREC}[TREC]{Text REtrieval Conference}
\acrodef{TRECVID}[TRECVID]{TREC Video Retrieval Evaluation}
\acrodef{TTL}[TTL]{Time-To-Live}
\acrodef{UCD}[UCD]{Use Case Diagram}
\acrodef{UDC}[UDC]{Universal Decimal Classification}
\acrodef{uGAP}[uGAP]{User-oriented Graded Average Precision}
\acrodef{UI}[UI]{User Interface}
\acrodef{UML}[UML]{Unified Modeling Language}
\acrodef{UMT}[UMT]{User Management Tool}
\acrodef{UMTS}[UMTS]{Universal Mobile Telecommunication System}
\acrodef{UoM}[UoM]{Utility-oriented Measurement}
\acrodef{UPC}[UPC]{Universal Product Code}
\acrodef{URI}[URI]{Uniform Resource Identifier}
\acrodef{URL}[URL]{Uniform Resource Locator}
\acrodef{URN}[URN]{Uniform Resource Name}
\acrodef{USM}[USM]{User Storing Manager}
\acrodef{VA}[VA]{Visual Analytics}
\acrodef{VAIRE}[VAIR\"{E}]{Visual Analytics for Information Retrieval Evaluation}
\acrodef{VATE}[VATE$^2$]{Visual Analytics Tool for Experimental Evaluation}
\acrodef{VIRTUE}[VIRTUE]{Visual Information Retrieval Tool for Upfront Evaluation}
\acrodef{VD}[VD]{Virtual Document}
\acrodef{VDM}[VDM]{Visual Data Mining}
\acrodef{VIAF}[VIAF]{Virtual International Authority File}
\acrodef{VL}[VL]{Visual Language}
\acrodef{VoIP}[VoIP]{Voice over IP}
\acrodef{VS}[VS]{Visual Sentence}
\acrodef{W3C}[W3C]{World Wide Web Consortium}
\acrodef{WAN}[WAN]{Wide Area Network}
\acrodef{WHO}[WHO]{World Health Organization}
\acrodef{WLAN}[WLAN]{Wireless \acs{LAN}}
\acrodef{WP}[WP]{Work Package}
\acrodef{WS}[WS]{Web Services}
\acrodef{WSD}[WSD]{Word Sense Disambiguation}
\acrodef{WSDL}[WSDL]{Web Services Description Language}
\acrodef{WWW}[WWW]{World Wide Web}
\acrodef{XMI}[XMI]{\acs{XML} Metadata Interchange}
\acrodef{XML}[XML]{eXtensible Markup Language}
\acrodef{XPath}[XPath]{XML Path Language}
\acrodef{XSL}[XSL]{eXtensible Stylesheet Language}
\acrodef{XSL-FO}[XSL-FO]{\acs{XSL} Formatting Objects}
\acrodef{XSLT}[XSLT]{\acs{XSL} Transformations}
\acrodef{YAGO}[YAGO]{Yet Another Great Ontology}
\acrodef{YASS}[YASS]{Yet Another Suffix Stripper}

\begin{abstract}

In this paper, we describe our participation in the TREC Health Misinformation Track 2020. We submitted $11$ runs to the Total Recall Task and $13$ runs to the Ad Hoc task. Our approach consists of $3$ steps: (1) we create an initial run with BM25 and RM3; (2) we estimate credibility and misinformation scores for the documents in the initial run; (3) we merge the relevance, credibility and misinformation scores to re-rank documents in the initial run. To estimate credibility scores, we implement a classifier which exploits features based on the content and the popularity of a document. To compute the misinformation score, we apply a stance detection approach with a pretrained Transformer language model. Finally, we use different approaches to merge scores: weighted average, the distance among score vectors and rank fusion.

\end{abstract}

\section{Introduction}
\label{sec:introduction}




The spread of fake news and misinformation has become a serious issue affecting society in many ways. It can, for instance, influence public opinion to promote a political candidate and substantially affect the final election outcome. It can also have deleterious effects on peoples' reputation, especially on social networks. When it comes to health, it can harm people by guiding them to take wrong decisions, which, in the worst cases, can lead people to them injure themselves. During the COVID-19 pandemic, we all witness how misinformation can have dangerous and have severe consequences on peoples' choices and lives.

In this context, \ac{IR} systems have a central position, as they can play an active role in contrasting the spread of misinformation \cite{10.1145/3006299.3006315}. Documents which are relevant, credible and \emph{incorrect} represent a serious threat for users, and should be discarded or presented very low in rankings \cite{10.1145/3121050.3121072}. Therefore, there is a need to design \ac{IR} systems that can promote relevant, credible and correct information over incorrect information.

This paper presents out attempt to address this challenging problem through our system submitted to the TREC Health Misinformation Track 2020. The Track offers $2$ tasks: the goal of the \emph{Total Recall} task is to identify all the documents which convey misinformation, while the goal of the \emph{Ad Hoc} task is to promote documents which convey relevant, correct and credible information.

We design a solution which consists of $3$ main steps. First, we generate an initial ranking of $1\,000$ documents with BM25 and RM3~\cite{LavrenkoAndCroft}. This initial ranking accounts for only relevance and serves solely as an initial filter to identify a smaller set of possibly relevant documents. The subsequent step involves the estimation of credibility and misinformation scores. We compute them in an independent way and adjust for documents retrieved in the initial run. 

To estimate the credibility score, we design a classifier which combines the predictions of $4$ different models. As features, we use both content features (i.e.~number of CSS definitions and readability) and social features (i.e.~features based on page rank). The classifier is trained on the Microsoft Credibility dataset~\cite{schwarz2011augmenting}.

To estimate the misinformation score, we use a popular approach from the fact checking domain. First, we use a stance detection \cite{mohammad2016semeval} model to identify the stance of a topic inside a document, then, we determine whether the stance agrees with the topic or not. As the stance detection model we use a Roberta~\cite{liu2019roberta} based classifier trained on the Fake News Challenge-1 dataset.\footnote{\url{http://www.fakenewschallenge.org/}}

As last step, we need to merge the relevance, credibility and misinformation scores to re-rank documents in the initial run. Since all scores result from different systems, we first normalize them as z\_scores. Then, we implement $3$ different strategies to merge those scores: (1) we consider the weighted average; (2) we consider the best score for each aspect and compute the distance between documents scores and the best scores; (3) we use reciprocal rank fusion~\cite{rrf}.

The paper is organized as follows: Section~\ref{sec:track_description} provides a brief description of the Health Misinformation Track; Section~\ref{sec:approach} describes our approach in details; Section~\ref{sec:experiments} reports the performance of our runs and discusses the experimental results; Section~\ref{sec:conclusions} presents conclusions and future work.

\section{Track Description}
\label{sec:track_description}

The TREC Health Misinformation Track 2020 includes two different tasks, the Total Recall task and the Ad Hoc task, presented in Section~\ref{subsec:task}. Both tasks use the CommonCrawl News crawl\footnote{\url{https://commoncrawl.org/2016/10/news-dataset-available/}}, which is described in Section~\ref{subsec:dataset}. An accurate description of the tasks can be found in the track overview paper~\cite{overview-misininfo-2020}.

\subsection{Tasks}
\label{subsec:task}

The goal of the Health Misinformation Track is to design and develop \ac{IR} Systems able not only to retrieve relevant documents, but also to rank credible and correct documents before not credible and incorrect documents. The Total Recall and Ad Hoc tasks are mirroring each other: the purpose of the Total Recall task is to retrieve all documents that are useful and incorrect, thus \emph{harmful} documents that convey misinformation; the purpose of the Ad Hoc task is to rank \emph{helpful} documents, i.e.~useful, credible, and correct documents, above harmful documents.

\subsection{Test Collection}
\label{subsec:dataset}

The corpus of the Total Recall and Ad Hoc Tasks consits of news articles in the CommonCrawl News crawl, sampled from January, 1st 2020 to April 30th, 2020. The dataset includes articles in different languages, but non English documents are considered not useful.

The tasks provide $49$ topics (officially $50$, but topics $7$ and $17$ are repeated). All topics are about health  and COVID-19. Each topic has a title, a description, which reformulates the title as a question, a yes/no answer, which is the actual answer to the description field based on the provided evidence, and a narrative, to describe helpful and harmful documents in relation to the given topic.

Documents were judged with respect to three aspects: \emph{usefulness}, i.e.~whether a document contains useful information to answer the question in the topic description field, \emph{credibility}, i.e.~whether a document is considered credible by the assessor, and \emph{correctness}, i.e.~whether a document answers to the topic question as specified in the answer field.


\section{Approach}
\label{sec:approach}

In this section, we present our approach for the Total Recall and Ad Hoc tasks. For both tasks, the challenge is to rank documents not solely by their usefulness score, but also by their credibility and correctness scores. We estimate usefulness with credibility, i.e.~we use basic \ac{IR} models to retrieve relevant documents. Similarly, we estimate correctness by computing the misinformation score with a fact checking approach.

Since it was not computationally feasible to estimate a relevance, credibility and misinformation score for each document in the collection, we break down the problem as follows:
\begin{enumerate}
    \item We exploit the entire collection and produce an initial run $R$ based on relevance only;
    \item We estimate credibility and misinformation scores for each document $d \in R$ and for each topic $q$;
    \item We re-rank documents in $R$ based on relevance, credibility and misinformation.
\end{enumerate}


In the following, Section~\ref{sec:relevance_ranker} presents how we build the the initial run based on relevance, Section~\ref{subsec:cedibility_ranker} describes how we compute credibility scores, Section~\ref{subsec:misinfo_ranking} details how we compute misinformation scores and finally Section~\ref{subsec:merging} explains how we merged relevance, credibility and misinformation scores to re-rank documents and produce the final run.

\subsection{Estimating Relevance}
\label{sec:relevance_ranker}


We select two simple unsupervised models to generate the initial run $R$: BM25 and a language model combined with pseudo-relevance feedback RM3~\cite{LavrenkoAndCroft}. To retrieve documents, we use the topic title and description as query. For each topic, we retrieve the top $1\,000$ documents, which became the candidate initial run $R$.

To create the index and implement the retrieval models, we use the already available implementations by Anserini\footnote{https://github.com/castorini/Anserini}, which is a java toolkit built over the open-source search library Lucene\footnote{https://lucene.apache.org}. To build the index, we remove standard English stop words\footnote{\href{https://github.com/castorini/anserini/blob/master/src/main/resources/solr/anserini/conf/stopwords_en.txt}{Stopwords list.}} and use Porter stemmer.

BM25 has two parameters $b$ and $k1$, which represent a correction factor for document length normalization and control term frequency saturation respectively. We set $b=0.4$ and $k1=0.9$, which is the default Anserini configuration. 
For RM3 model, we also use the default Anserini configuration, which is $fb_{terms}=10, fb_{docs}=10$ and $\text{original\_query\_weight}= 0.5$.

\subsection{Estimating Credibility}
\label{subsec:cedibility_ranker}

We frame the credibility estimation prediction as a classification problem. 
Formally, consider a fixed set of credibility classes $L = \{l_{1}, l_{2},\dots, l_{n}\}$, representing the credibility labels for a document, for example $L = \{1, 2, 3, 4, 5\}$, where the higher the integer, the higher the credibility. Each document $d$ is represented by a feature vector $\vec{x} \in X$. Given a training set of $m$ labeled documents $\{(\vec{x}_{1},l_{1}), \dots, (\vec{x}_{m},l_{m})\}$, where $\vec{x}_i$ represents the feature vector of the $i^{th}$ training document $d_{i}$ and $l_{i}$ its class label, the classifier fits a function $\gamma: X\rightarrow L$, such that for a new unseen document $d_{k}$ and its features $\vec{x}_{k}$, the classifier can predict $l_{k}\in L$.

We consider credibility as a topic independent property of documents, meaning that its estimation can be based solely on the document, disregarding which topic the document was retrieved for. Therefore, we propose a classification algorithm which approximates the credibility of a document based on its textual and design content, and publicly available information estimating documents popularity.

\subsubsection*{Credibility features} 

Inspired by~\citet{10.1007/978-3-642-36973-5_47}, we devise our feature set to represent documents. These features are summarized in Table~\ref{tab:credibility-features} and organized into two groups as suggested in~\citep{10.1007/978-3-642-36973-5_47}: Content Features and Social Features.
\begin{description}
\item[Content Features] are extracted from the raw HTML content of the document. They consist of features based on the Web page textual content (i.e.~text readability), or based on the Web page structure (i.e.~number of CSS definitions);

\item[Social Features] Publicly available information about the Web page can be a good indicator of its credibility. For instance, the popularity of a document can be approximated by its PageRank. The underlying assumption is that, incoming links to a Web page can be seen as endorsements for the Web page and, thus, they could be a good indicator for a Web page popularity~\cite{10.1007/978-3-642-36973-5_47}. We assume that popular pages are also reliable.
\end{description}

\begin{table}[h]
\small
\centering
\caption{Set of features used for credibility detection.}
\label{tab:credibility-features}
\begin{tabular}{@{}ll@{}}
\toprule
\multicolumn{2}{c}{\textbf{Content Features}} \\ \midrule
css\_definitions & \# of CSS definitions in the raw HTML content of the document \\
text\_readability & Estimated school grade level required to understand the text extracted using textstat\tablefootnote{\url{https://pypi.org/project/textstat/}}\\ \midrule
\multicolumn{2}{c}{\textbf{Social Features}} \\ \midrule
pr\_rank & The rank position of a document based on page rank extracted using OpenPageRank API\tablefootnote{\url{https://www.domcop.com/openpagerank/}} \\
page\_rank\_integer & The rounded page rank score of the document extracted using OpenPageRank API \\
page\_rank\_decimal & The page rank score of the document extracted using OpenPageRank API \\
toplevel\_domain &  Check weather the Web page URL contains .edu, .gov, .com, etc.\\
\bottomrule
\end{tabular}
\end{table}

\subsubsection*{Classification Models} 

As the credibility model, we use four different state-of-the-art classification approaches implemented with the library scikit-learn~\footnote{\url{https://scikit-learn.org/stable/}}: Logistic Regression, Random Forest, SVM Logistic, and Naive Bayes. We combine the predictions of the classifiers using a soft VotingClassifier, which predicts the class label based on the argmax of the sums of the predicted probabilities. For all the different classification approaches we used the default parameters of the scikit-learn library.

\subsubsection*{Training}

Since there are no credibility assessments for the Common Crawl News sample used for this track, we could not train our model on this corpus. We instead use the well known Microsoft Credibility dataset~\cite{schwarz2011augmenting}, which is a Web dataset not only containing health-related credibility assessments, but also more general topics (e.g., celebrities, politics). This dataset consists of a set of Web pages retrieved for different queries. For each retrieved Web page, it contains its URL, rank position, and a credibility label in $L = \{1,2,3,4,5\}$. 

To simplify the task, we consider a binary classifier. Therefore, we map credibility labels to two classes as follows: labels $1,2,3$ are mapped to $0$ and $4,5$ are mapped to 1.

\subsection{Estimating Misinformation}
\label{subsec:misinfo_ranking}

We frame the estimation of misinformation as a stance detection problem. The goal of stance detection is: given a claim $C$ and a statement $S$, determine if $S$ agrees with, disagrees with, or is neutral towards $C$. In this, stance detection is generally approached as a supervised learning problem with labelled pairs ($C$,$S$)~\cite{mohammad2016semeval,augenstein2016stance}, and is similar to the problem setting in other fact checking tasks e.g. FEVER~\cite{thorne2018fever}. 


We start with the set of topics (queries) and answers provided for the task $\{(t_{1}, a_{1}), (t_{2}, a_{2}), ..., (t_{50}, t_{50})\} \in M$, with $a_{k} \in \{0, 1\}$ corresponding to ``no'' and ``yes,'' and corpus of articles after the initial run $R$. 
Each $(t_{k}, a_{k})$ tuple is paired with a subset of documents $\hat{R}_{k} \subset R$ using one of the ranking approaches described in \S\ref{sec:relevance_ranker}. The goal is then to predict the stance of each document $d_{k}^{(i)} \in \hat{R}_{k}$ towards the topic $t_{k}$, and then rank each document based on the relationship of the predicted stance towards the answer $a_{k}$.

More concretely, a stance prediction model takes as input the topic $t_{k}$ and a document $d_{k}^{(i)}$, and produces probabilities $\mathbb{P}(l|t_{k}, d_{k}^{(i)})$ for labels $l \in \{0, 1, 2\}$ corresponding to levels of disagreement, agreement, and neutrality. The final misinformation score $s$ then takes into account how much the model predicts the document both disagrees and agrees with the topic via a subtractive measure.
\begin{equation}
    s(k, i) = \mathbb{P}\left(1 - a_{k}|t_{k}, d_{k}^{(i)}\right) - \mathbb{P}\left(a_{k}|t_{k}, d_{k}^{(i)}\right)
\end{equation}

The effect of this scoring function is as follows: if the answer of a given topic is ``no'', then the score will take the probability that the article \textit{agrees} with the topic and subtract from this the probability that it \textit{disagrees} with the topic. If the answer of a given topic is ``yes'', then the score will take the probability that the article \textit{disagrees} with the topic and subtract from this the probability that it \textit{agrees} with the topic. The purpose of using the subtractive measure is to take into consideration articles which are neutral towards the topic. The net effect is that articles which are more likely to disagree with the answer will have higher scores than articles which are more likely to agree with the answer.

\subsubsection*{Training}
We use a large pretrained transformer language model as our stance detection model, namely Roberta~\cite{liu2019roberta} from the HuggingFace transformers library~\cite{Wolf2019HuggingFacesTS}. Roberta is a large transformer model pretrained on massive unlabelled text corpora using a masked language modeling objective. Models trained in this manner require only to be fine-tuned on downstream tasks, and have shown to achieve state of the art performance after doing so~\cite{devlin2019bert}. In this, we only require a labelled dataset on which to fine-tune the model. 

As no labelled dataset currently exists for the specific problem of health related stance detection within the Common Crawl News corpus, a large general domain dataset is used to train the stance detection model. In this, we use the Fake News Challenge-1 dataset\footnote{\url{https://github.com/FakeNewsChallenge/fnc-1}}, which consists of full text news articles, headlines, and stance labels between them. The labels can be one of \{unrelated, discuss, agree, disagree\}. This dataset is unique in that it is designed for performing stance detection on full text articles, making it a good fit as a source of training data. We follow previous best practice when working with this dataset and train a separate model for the binary task of predicting whether or not the headline is related to the article, and a separate model for predicting the stance, achieving comparable results to previous work on the specific task of stance detection (73.5 macro F1 score)~\cite{slovikovskaya2020transfer}.

Prior to ranking articles using the stance prediction model, we preprocess the topics by manually converting each topic description from a question to a statement in order to better match the style of data the model was trained on. In addition, the articles in the dataset tend to have a large amount of irrelevant information as a result of imperfect cleaning of the data, while the Roberta model has a limited token capacity of 512, meaning we can only perform prediction using subsets of large articles. Given this, as a preprocessing step we filter the article content by locating the first occurrence of the topic string in the description e.g. ``ibuprofen,'' and perform misinformation ranking on the article starting from the sentence containing this string.

\subsection{Merging Scores}
\label{subsec:merging}

In this section, we explain how we re-rank documents by considering relevance, credibility and misinformation scores. 
Observe that all the scores result as output of different models, thus, they are not directly comparable. 
Therefore we standardize the scores by computing per topic z-scores as follows:
\begin{equation}
    z\_score_{a}(d, t) = \frac{x_{a}(d, t) - \mu_{a}(t)}{\sigma_{a}(t)}
\end{equation}
where $a\in A$ refers to the aspect $A=\{\textit{relevance}, \textit{credibility}, \textit{misinformation}\}$, $x_{a}(d, t)$ is the relevance, credibility or misinformation score of document $d$ for topic $t$, $\mu_{a}(t)$ and $\sigma_{a}(t)$ are the mean and standard deviation of relevance, credibility or misinformation scores for all documents retrieved for topic $t$.

We use three different approaches to combine relevance, credibility and misinformation scores in a single score: weighted average, distance from the best and ranking fusion. The final single score is used to re-rank documents and produce the final run which accounts for all the three aspects.

\subsubsection*{Weighted Average}
For a fixed topic $t$, we compute the weighted sum of z-scores for each document $d\in R$ as follows: 
\begin{equation}
    S(d, t) = \sum_{a \in A} w_{a} \times z\_score_{a}(d, t) 
\end{equation}
where $w_{a}$ is the weight associated to aspect $a \in A$ (e.g.~weight given to relevance score). 
We arbitrarily set the weights $w_{a}$ to be uniform or we assign more weight to one aspect over the others, depending on the run.

\subsubsection*{Distance from the Best Score}

Given a topic $t$, we consider the highest z\_score across aspects as a reference score. More formally, let $z\_score_{a}^\star(t) = \max\{z\_score_{a}^\star(d, t) | d \in R\}$ be the highest z\_score for a given topic and aspect. We define the \emph{best score vector} as the vector whose coordinates are $z\_score_{a}^\star(t)$:
\begin{equation}
    z\_score^\star(t) = [z\_score_{rel}^\star(t), z\_score_{cred}^\star(t), z\_score_{mis}^\star(t)]
    \label{eq:max_score}
\end{equation}

Similarly, each document can be represented as a vector of z\_scores. Therefore, we define the final ordering of documents by computing the distance of each document vector from the best score vector: 
\begin{equation}
    S(d, t) = dist(z\_score^\star(t), z\_score(d,t))
\end{equation}
where $z\_score(d,t)$ denotes the score vector for document $d$ with respect to topic $t$.
The closer the document vector to the best score vector, the higher a document should be in the ranking. We instantiate $dist$ with the well-known distances, Euclidean and Chebyshev.

Observe that Equation~\eqref{eq:max_score} can be modified by taking the minimum instead of the maximum for certain aspects. For example, for the Total Recall task the purpose is to find documents that disagree with the answer, thus we take the maximum of the misinformation score, while for the Ad Hoc task the purpose is to find documents that agree with the answer, thus we take the minimum of the misinformation score.





\subsubsection*{Rank Fusion}

We consider $3$ different runs  $R_{a}$ by ranking documents solely based on a given aspect $a$. We merge these $3$ runs into a single runs with Reciprocal Rank Fusion (RRF)~\cite{rrf}. Specifically, for each topic $t$, we compute the fused score of each document as follows:
\begin{equation}
    RRF\_score(d, t) = \sum_{a \in A}\frac{1}{k+\text{rank}_{a}(d, t)}
\end{equation}
where $\text{rank}_{a}(d, t)$ is the rank position of document $d$, when ranked with respect to topic $t$ and aspect $a$; $k$ is a parameter to control the impact of bad ranking systems, which can retrieve bad documents at high positions in the ranking. We set $k=60$, the value used in the original paper by~\citet{rrf}.


\section{Experimental Evaluation}
\label{sec:experiments}

\subsection{Submitted Runs}
\label{subsec:submission}

\subsubsection*{Total Recall Task}

We submitted a total of $11$ runs to the Total Recall Task, described in Table~\ref{tab:submission_total_recall}. We combine different cut-offs for re-ranking and different merging approaches. We also create some runs by exploiting just the credibility or misinformation scores, specifically \texttt{run4}, \texttt{run5} and \texttt{run9}.

Observe that for all runs, we used the reversed credibility (i.e.~by multiplying the score $\times-1$), so that low credibility documents are placed towards the beginning of the ranking. Indeed, for the Total Recall task we want the documents with low credibility, but high misinformation and relevance score, to be retrieved at the top of the ranking.

\begin{table}[htb]
\centering
\small
\caption{Runs submitted to the Total Recall Task.}
\label{tab:submission_total_recall}
\resizebox{\textwidth}{!}{%
\begin{tabular}{@{}lll@{}}
\toprule
RUNID & Initial Run & Re-ranking \\ \midrule
\texttt{run1}    &   BM25\_description & top $200$ documents using weighted average $w_{rel}=-w_{cred}=w_{mis}$ \\
\midrule
\texttt{run2} & RM3\_description & top $100$ documents using weighted average $-w_{cred}=w_{mis}$\\
\texttt{run3} & RM3\_description & top $100$ documents using weighted average $-w_{rel}=-w_{cred}=w_{mis}$\\
\midrule
\texttt{run4} & RM3\_description & top $100$ documents using only the reversed credibility score\\
\texttt{run5} & RM3\_description & top $100$ documents using only the misinformation score \\
\midrule
\texttt{run6} & - & Reciprocal rank fusion of \texttt{run4} and \texttt{run5} \\
\midrule
\texttt{run7} & RM3\_description & top $100$ documents using the Euclidean distance from the best score (min cre score)\\
\texttt{run8} & RM3\_description & top $100$ documents using the Chebyshev distance from the best score (min cre score)\\
\midrule
\texttt{run9} & RM3\_description & top $200$ documents using only the reversed credibility score\\
\midrule
\texttt{run10} & RM3\_description & top $300$ documents using the Euclidean distance from the best score (min cre score)\\
\midrule
\texttt{run11}  & - & Reciprocal rank fusion of all runs \\
\bottomrule
\end{tabular}
}
\end{table}

\subsubsection*{Ad Hoc Task}

We submitted a total of $13$ runs to the Ad Hoc Task, described in Table~\ref{tab:submission_ad_hoc}. First, we submitted $2$ baseline runs: \texttt{adhoc\_run1} and \texttt{adhoc\_run2}. These can be seen as runs where documents are sorted just by their relevance score. As for the Total Recall Task, we submitted some runs that exploit just the credibility or misinformation score, \texttt{adhoc\_run9}, \texttt{adhoc\_run10}, \texttt{adhoc\_run12}, \texttt{adhoc\_run13}.  Furthermore, we combine different cut-offs for re-ranking and different merging approaches.

Observe that we use the reversed misinformation score, since a high misinformation score means a high disagreement between the document and the topic answer. Therefore, those documents should be placed close to the bottom of the ranking.

\begin{table}[htb]
\centering
\caption{Runs submitted to the Ad Hoc Task.}
\label{tab:submission_ad_hoc}
\resizebox{\textwidth}{!}{%
\begin{tabular}{@{}lll@{}}
\toprule
RUNID & Initial Run & Re-ranking \\ \midrule
\texttt{adhoc\_run1}    &  BM25\_description & baseline, no re-ranking            \\
\texttt{adhoc\_run2}   &  RM3\_description & baseline, no re-ranking           \\
\midrule
\texttt{adhoc\_run3}    &  BM25\_description & top $100$ documents using weighted average $w_{rel}=w_{cred}=-w_{mis}$         \\
\texttt{adhoc\_run4}    &  RM3\_description & top $100$ documents using weighted average $w_{rel}=w_{cred}=-w_{mis}$          \\
\texttt{adhoc\_run5}    &  BM25\_description & top $100$ documents using weighted average $w_{rel}=0.6$ and $w_{cred}=-w_{mis}=0.2$          \\
\texttt{adhoc\_run6}    &   RM3\_description & top $100$ documents using weighted average $w_{rel}=0.6$ and $w_{cred}=-w_{mis}=0.2$     \\
\midrule
\texttt{adhoc\_run7}   &  RM3\_description & top $100$ documents using the Euclidean distance from the best score (min mis score)          \\
\texttt{adhoc\_run8}    &   RM3\_description & top $100$ documents using the Chebyshev distance from the best score (min mis score)         \\
\midrule
\texttt{adhoc\_run9}    &   RM3\_description & top $100$ documents using only the credibility score
          \\
\texttt{adhoc\_run10}  &   RM3\_description & top $100$ documents using only the reversed misinformation score           \\
\midrule
\texttt{adhoc\_run11}   & -  & Reciprocal rank fusion of \texttt{adhoc\_run1}, \texttt{adhoc\_run9} and \texttt{adhoc\_run10}           \\
\midrule
\texttt{adhoc\_run12}  &   RM3\_description & top $200$ documents using only the credibility score          \\
\texttt{adhoc\_run13}   &    RM3\_description & top $200$ documents using only the reversed misinformation score
          \\ \bottomrule
\end{tabular}
}
\end{table}

\subsection{Results}
\label{subsec:results}

\subsubsection*{Total Recall Task}

\begin{table}[htb]
\centering
\caption{Total Recall Task: our runs compared to the overall median results of submitted runs.}
\label{tab:recall}
\begin{tabular}{@{}ll@{}}
\toprule
\textbf{RUNID} & \textbf{Rprec} \\ \midrule
\texttt{run9} & 0.0866 \\
\texttt{run1} & 0.0936 \\
median (TREC) & 0.0976 \\
\texttt{run4} & 0.1090 \\
\texttt{run3} & 0.1115 \\
\texttt{run10} & 0.1142 \\
\texttt{run6} & 0.1173 \\
\texttt{run8} & 0.1226 \\
\texttt{run2} & 0.1237 \\
\texttt{run7} & 0.1267 \\
\texttt{run11} & 0.1274 \\
\texttt{run5} & 0.1303 \\ \bottomrule
\end{tabular}
\end{table}

Table~\ref{tab:recall} reports Rprec scores for our runs submitted at the Total Recall Task: $9$ out of $11$ runs are above the task median. However, Rprec scores are overall quite low, our best score is $0.1303$, showing that finding useful but not correct documents is a challenging task. 

Our best run is \texttt{run5}, which re-rank documents solely with the misinformation score. This is a promising results, suggesting that our approach to estimate misinformation goes towards the correct direction. Among the approaches to merge scores, rank fusion seems the most effective, followed by the Euclidean distance from the best score.

Our worst run is \texttt{run9}, which re-rank the first $200$ documents by their reversed credibility score. Clearly, this strategy is too harsh, since it places close to the top of the ranking documents that are not relevant, thus not credible and not correct. The second-to-last run is \texttt{run1}, which is below the task median. This run re-ranks the top $200$ documents by merging the scores with a weighted average with uniform weights. We hypothesise that this is due to the reversed credibility score, indeed \texttt{run4} re-ranks documents solely with the reversed credibility score and is the third-to-last run, just above the task median. It might be that a low credibility score denotes very low quality documents, which represent spam, and thus are not relevant.

\subsubsection*{Ad Hoc Task}

\begin{table}[b]
\centering
\caption{Auxiliary Table with mapping of qrels and measures used in Table~\ref{tab:adhoc}.}
\label{tab:mapping}
\begin{tabular}{@{}ccc@{}}
\toprule
\textbf{MappingID} & \textbf{Qrels} & \textbf{Measure} \\ \midrule
0 & 2aspects.correct-credible & cam\_map \\
1 & 2aspects.useful-credible & cam\_map \\
2 & 3aspects & cam\_map\_three \\
3 & binary.useful & ndcg \\
4 & binary.useful-correct & ndcg \\
5 & binary.useful-correct-credible & ndcg \\
6 & binary.useful-credible & ndcg \\
7 & graded.harmful-only & compatibility \\
8 & graded.helpful-only & compatibility \\ \bottomrule
\end{tabular}%
\end{table}

\begin{table}[htb]
\centering
\caption{Ad Hoc Task: our runs compared to the overall median results of submitted runs. For 7 the lower the score the better.}
\label{tab:adhoc}
\begin{tabular}{@{}llllllllll@{}}
\toprule
\textbf{RUNID} & \multicolumn{1}{c}{\textbf{0}} & \multicolumn{1}{c}{\textbf{1}} & \multicolumn{1}{c}{\textbf{2}} & \multicolumn{1}{c}{\textbf{3}} & \multicolumn{1}{c}{\textbf{4}} & \multicolumn{1}{c}{\textbf{5}} & \multicolumn{1}{c}{\textbf{6}} & \multicolumn{1}{c}{\textbf{7}} & \multicolumn{1}{c}{\textbf{8}} \\ \midrule
\texttt{adhoc\_run1} & 0.1911 & 0.2765 & 0.2369 & 0.6077 & 0.4997 & 0.4853 & 0.5774 & 0.1197 & 0.3679 \\
\texttt{adhoc\_run10} & 0.2084 & 0.3088 & 0.2631 & 0.6027 & 0.4841 & 0.4702 & 0.5632 & 0.1263 & 0.3736 \\
\texttt{adhoc\_run11} & 0.2105 & 0.3354 & 0.2823 & \textbf{0.6333} & 0.4945 & 0.4771 & \textbf{0.5970} & 0.1456 & 0.3985 \\
\texttt{adhoc\_run12} & 0.1558 & 0.2563 & 0.2124 & 0.5675 & 0.4401 & 0.4416 & 0.5573 & \textbf{0.0826} & 0.3391 \\
\texttt{adhoc\_run13} & 0.2103 & 0.2935 & 0.2491 & 0.5926 & 0.4791 & 0.4936 & 0.5720 & 0.1092 & 0.3869 \\
\texttt{adhoc\_run2} & \textbf{0.2155} & 0.3318 & 0.2797 & 0.6163 & 0.4836 & 0.4769 & 0.5868 & 0.1446 & 0.4074 \\
\texttt{adhoc\_run3} & 0.2070 & 0.2924 & 0.2495 & 0.6170 & \textbf{0.5081} & \textbf{0.5065} & 0.5948 & 0.1208 & 0.4012 \\
\texttt{adhoc\_run4} & 0.2097 & 0.3302 & 0.2781 & 0.6229 & 0.4828 & 0.4690 & 0.5879 & 0.1385 & 0.4098 \\
\texttt{adhoc\_run5} & 0.1994 & 0.2919 & 0.2488 & 0.6187 & 0.5053 & 0.4961 & 0.5924 & 0.1247 & 0.3981 \\
\texttt{adhoc\_run6} & 0.2128 & \textbf{0.3389} & \textbf{0.2843} & 0.6223 & 0.4812 & 0.4723 & 0.5941 & 0.1487 & \textbf{0.4143} \\
\texttt{adhoc\_run7} & 0.2086 & 0.3297 & 0.2777 & 0.6227 & 0.4813 & 0.4656 & 0.5891 & 0.1355 & 0.4023 \\
\texttt{adhoc\_run8} & 0.2012 & 0.3241 & 0.2722 & 0.6189 & 0.4749 & 0.4569 & 0.5841 & 0.1338 & 0.3846 \\
\texttt{adhoc\_run9} & 0.1761 & 0.3008 & 0.2491 & 0.5859 & 0.4507 & 0.4410 & 0.5756 & 0.0893 & 0.3473 \\
median (TREC) & 0.1003 & 0.1717 & 0.1389 & 0.4699 & 0.338 & 0.3308 & 0.4471 & 0.0747 & 0.3337 \\ \bottomrule
\end{tabular}
\end{table}

Table~\ref{tab:mapping} and Table~\ref{tab:adhoc} report the results for our runs in the Ad Hoc task. For all measures, except for harmfulness, our runs are above the task median. On the other side, for harmfulness, all our runs are worse than the task median, meaning that we need to improve our strategy to move harmful documents towards the end of the ranking. 

Table~\ref{tab:adhoc} shows that our best runs always improve over our internal baselines (\texttt{adhoc\_run1} and \texttt{adhoc\_run2}). This is always true, except for the case of $2$ aspects, correctness and credibility, evaluated with \ac{CAM} instantiated with \ac{AP} (Table~\ref{tab:adhoc}, column $1$, $\text{id}=0$).

\begin{figure}[htbp]
\centering
\begin{subfigure}{.49\textwidth}
  \centering
  \includegraphics[width=\linewidth]{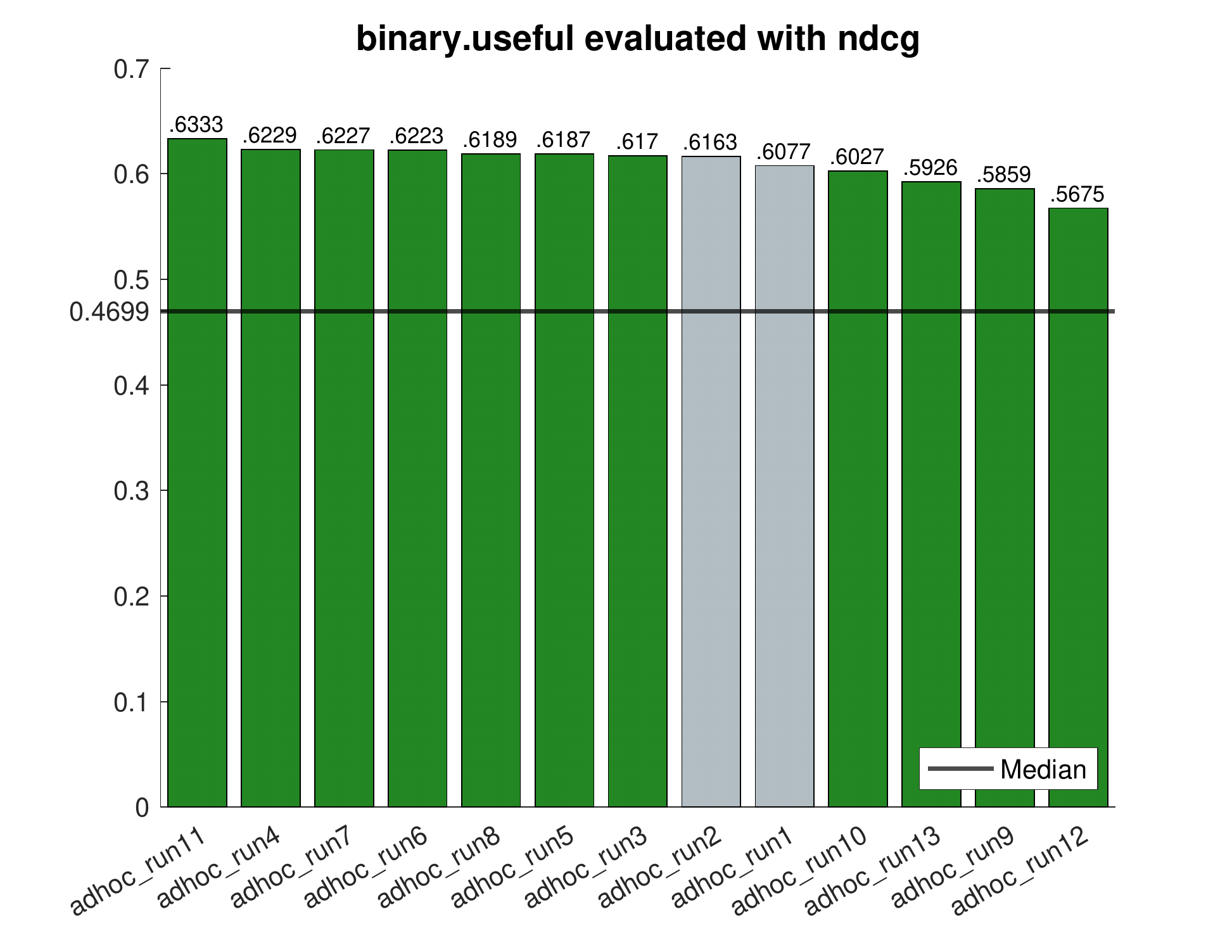}
  \caption{Only Useful, $\text{id}=3$}
  \label{fig:binary_useful}
\end{subfigure}
\begin{subfigure}{.49\textwidth}
  \centering
  \includegraphics[width=\linewidth]{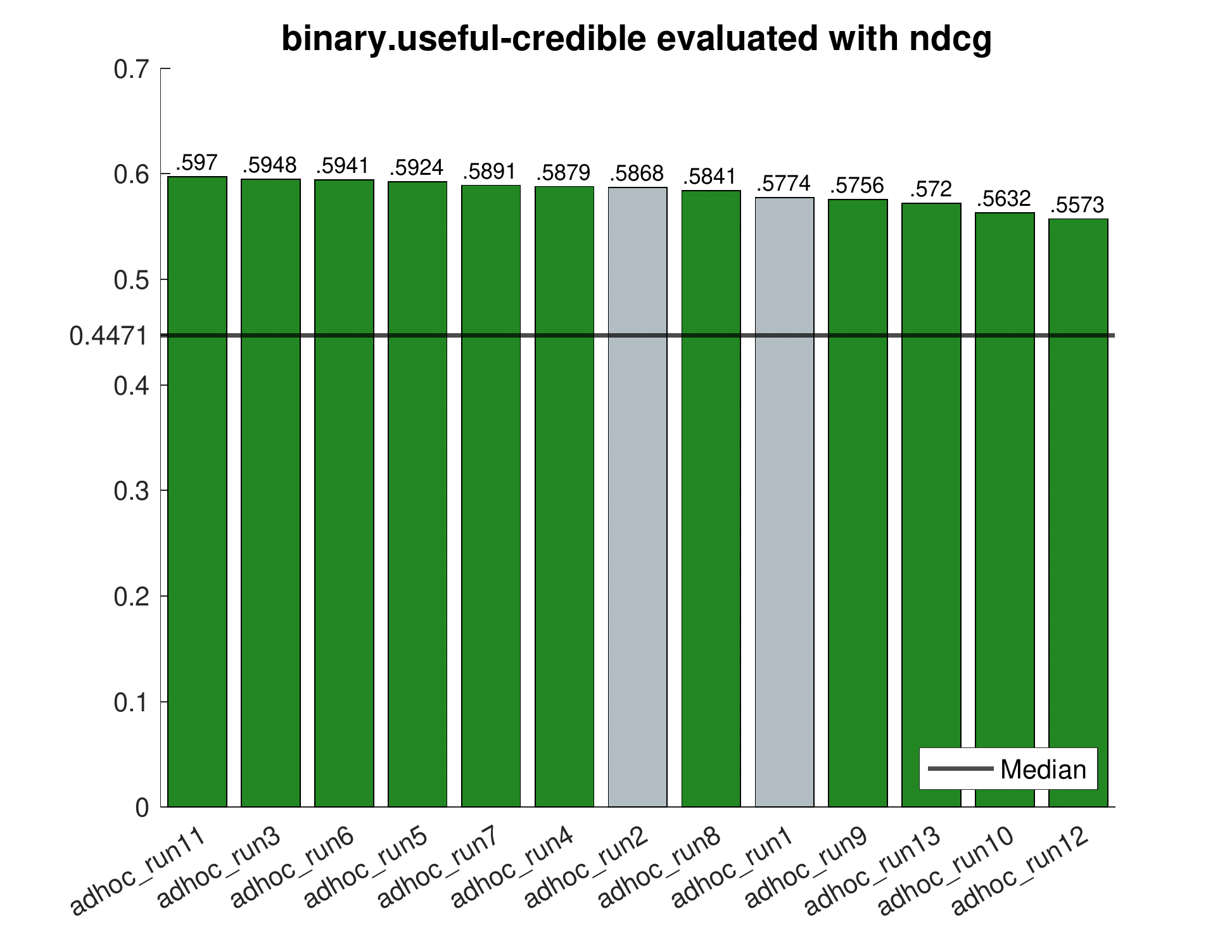}
  \caption{Useful and Credible, $\text{id}=6$}
  \label{fig:binary_useful_credible}
\end{subfigure}
\begin{subfigure}{.49\textwidth}
  \centering
  \includegraphics[width=\linewidth]{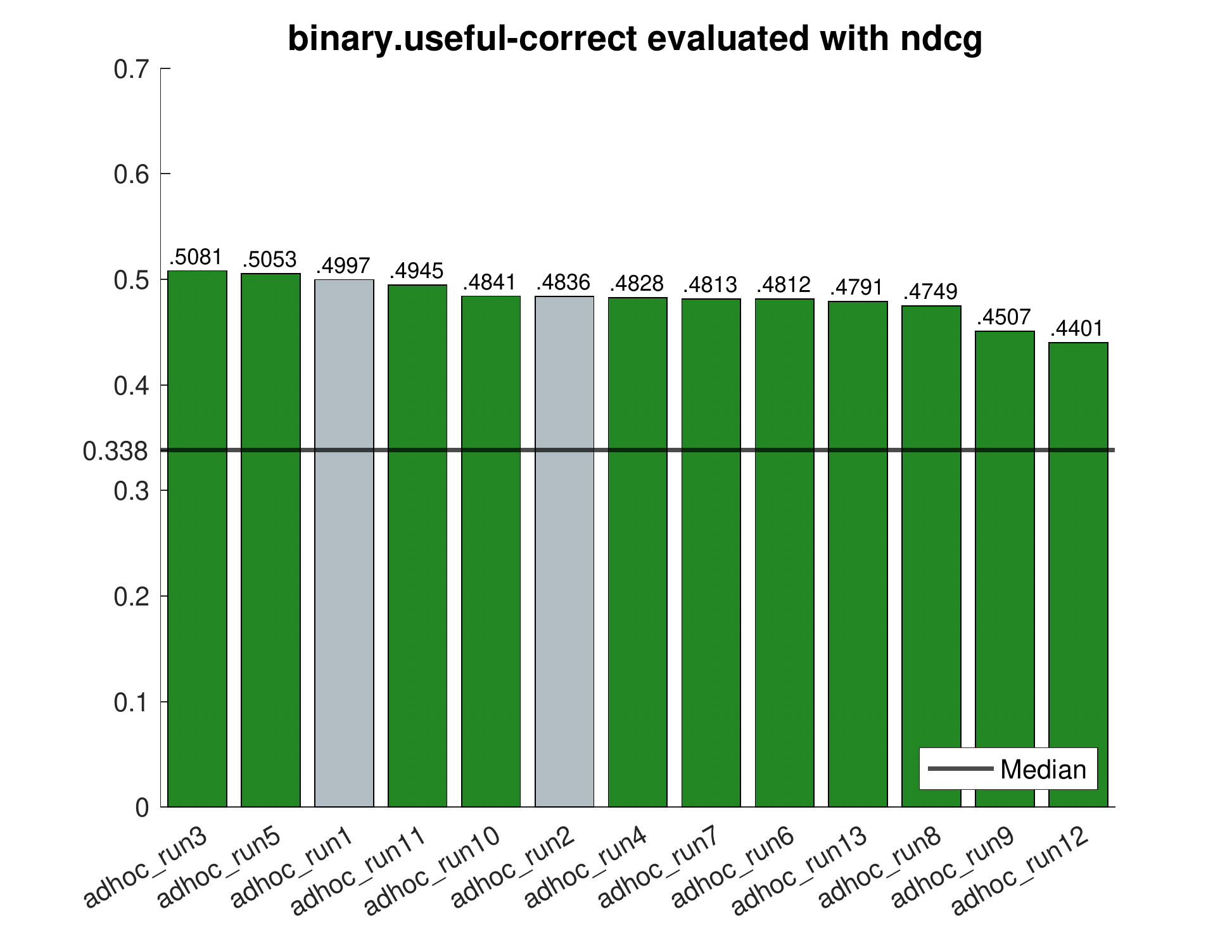}
  \caption{Useful and Correct, $\text{id}=4$}
  \label{fig:binary_useful_correct}
\end{subfigure}
\begin{subfigure}{.49\textwidth}
  \centering
  \includegraphics[width=\linewidth]{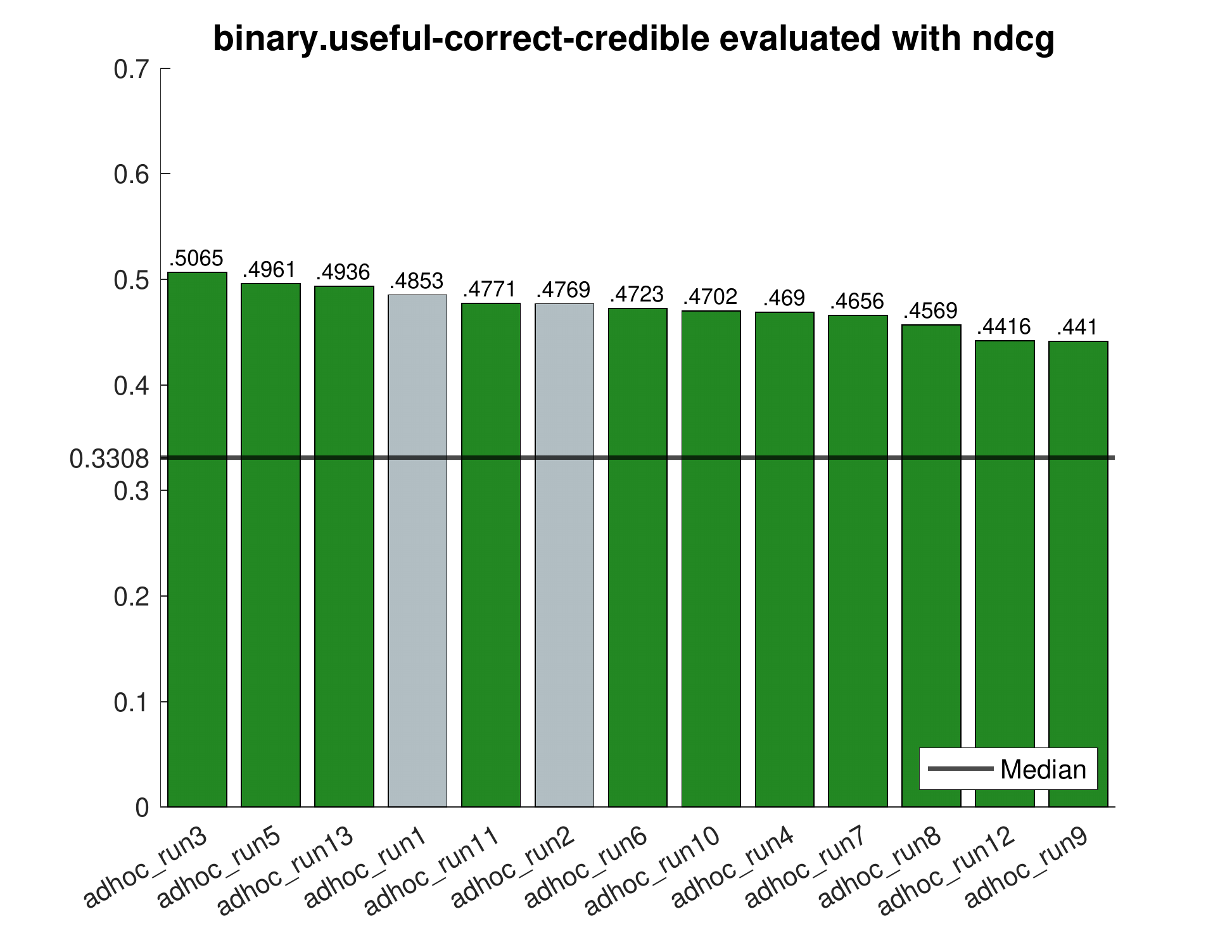}
  \caption{Useful, Correct and Credible, $\text{id}=5$}
  \label{fig:binary_useful_correct_credible}
\end{subfigure}
\caption{Ad Hoc Task: Our runs evaluated with \acs{nDCG} by mapping multi-aspect labels to binary labels.}
\label{fig:binary_evaluation}
\end{figure}

Figure~\ref{fig:binary_evaluation} shows the performance of our runs when evaluated with \acf{nDCG} with respect to binary multi-aspect labels: for example, in Figure~\ref{fig:binary_useful_credible} a document is mapped to $1$ if it is useful and credible, independently of its correctness. 

For usefulness in Figure~\ref{fig:binary_useful}, the best run is \texttt{adhoc\_run11}, which uses RRF with relevance, credibility and the reversed misinformation score. This is the best run also for usefulness and credibility in Figure~\ref{fig:binary_useful_credible}. Therefore, as shown by the Total Recall Task, RRF represents an effective approach to merge scores from different aspects.

Aggregating the scores with a weighted average is also quite effective: \texttt{adhoc\_run3} aggregates relevance, credibility and misinformation scores with uniform weights. This is the best run with respect to usefulness and correctness in Figure~\ref{fig:binary_useful_correct}, usefulness, correctness and credibility in Figure~\ref{fig:binary_useful_correct_credible}, and it is the second run with respect to usefulness and credibility in Figure~\ref{fig:binary_useful_credible}. Similarly, \texttt{adhoc\_run4}, which is computed as \texttt{adhoc\_run3} with RM3 instead of BM25, is the second best run with respect to usefulness in Figure~\ref{fig:binary_useful}.

\texttt{Adhoc\_run5} is the second best run for usefulness-correctness, and usefulness-correctness-credibility, in Figure~\ref{fig:binary_useful_correct} and Figure~\ref{fig:binary_useful_correct_credible} respectively. This run also exploits a weighted average with BM25, where more weight is assigned to relevance ($0.6$) and less to credibility and reversed misinformation ($0.2$).

On the other side, all the worst runs exploit a single aspect, instead of applying a merging strategy. \texttt{Adhoc\_run12} is the worst run with respect to usefulness in Figure~\ref{fig:binary_useful}, usefulness and credibility in Figure~\ref{fig:binary_useful_credible}, usefulness and correctness in Figure~\ref{fig:binary_useful_correct}, and is the second-to-last run with respect to all aspects in Figure~\ref{fig:binary_useful_correct_credible}. Similarly, \texttt{adhoc\_run9} is the worst or the second-to-last run in Figures~\ref{fig:binary_useful_correct_credible},~\ref{fig:binary_useful} and~\ref{fig:binary_useful_correct}. Both these runs simply re-rank the top $200$ and top $100$ documents just with the credibility score. Something analogous happens with \texttt{adhoc\_run13}, which re-ranks documents just with the reversed misinformation score. This is among the worst runs for usefulness, usefulness-credibility, usefulness-correctness in Figures~\ref{fig:binary_useful},~\ref{fig:binary_useful_credible} and~\ref{fig:binary_useful_correct}. This results suggest that if we want to optimize a run for relevance, credibility and correctness we need to use all $3$ scores. 

\begin{figure}[t]
\centering
\begin{subfigure}{.49\textwidth}
  \centering
  \includegraphics[width=\linewidth]{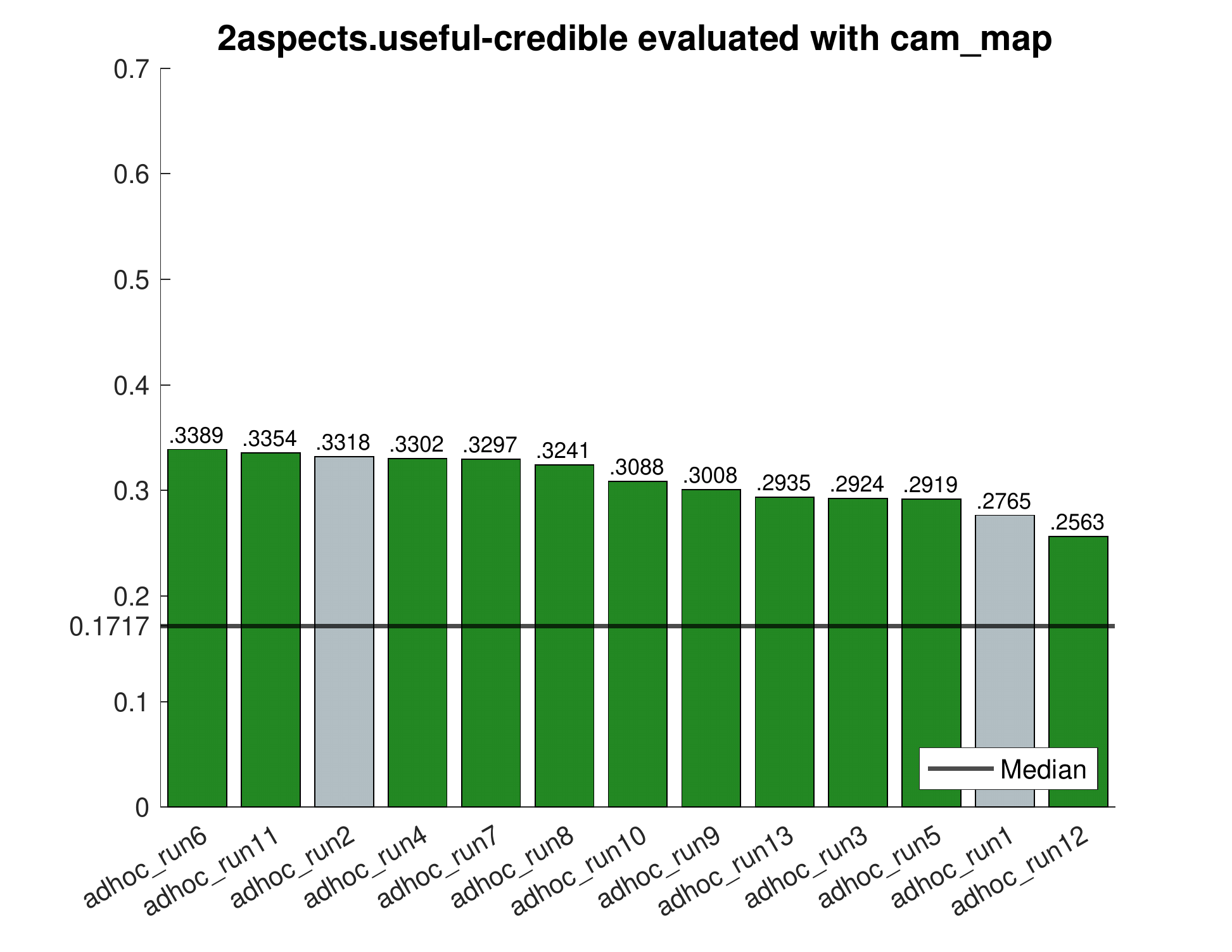}
  \caption{Useful and Credible, $\text{id}=1$}
  \label{fig:cam_useful_credible}
\end{subfigure}
\begin{subfigure}{.49\textwidth}
  \centering
  \includegraphics[width=\linewidth]{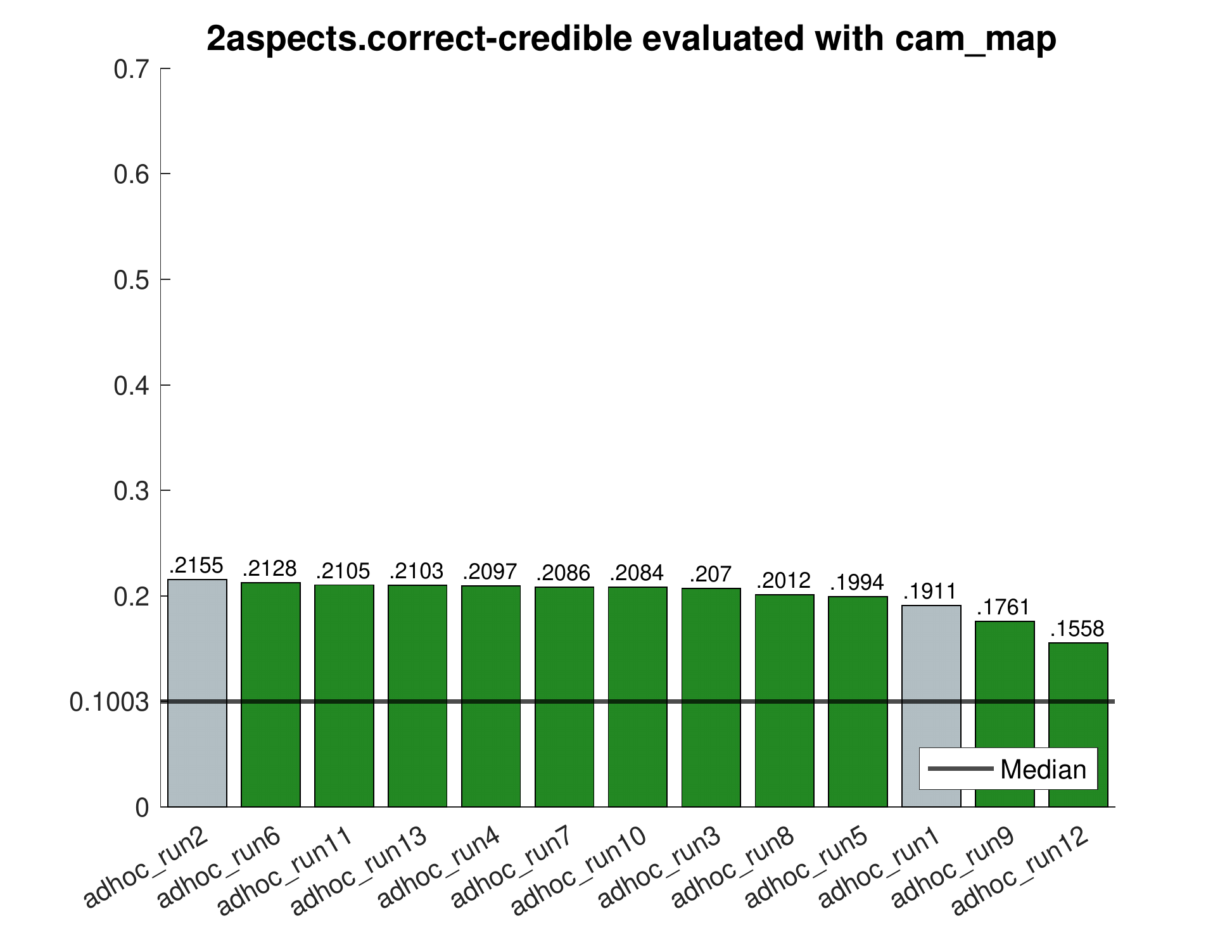}
  \caption{Correct and Credible, $\text{id}=0$}
  \label{fig:cam_correct_credible}
\end{subfigure}
\begin{subfigure}{.49\textwidth}
  \centering
  \includegraphics[width=\linewidth]{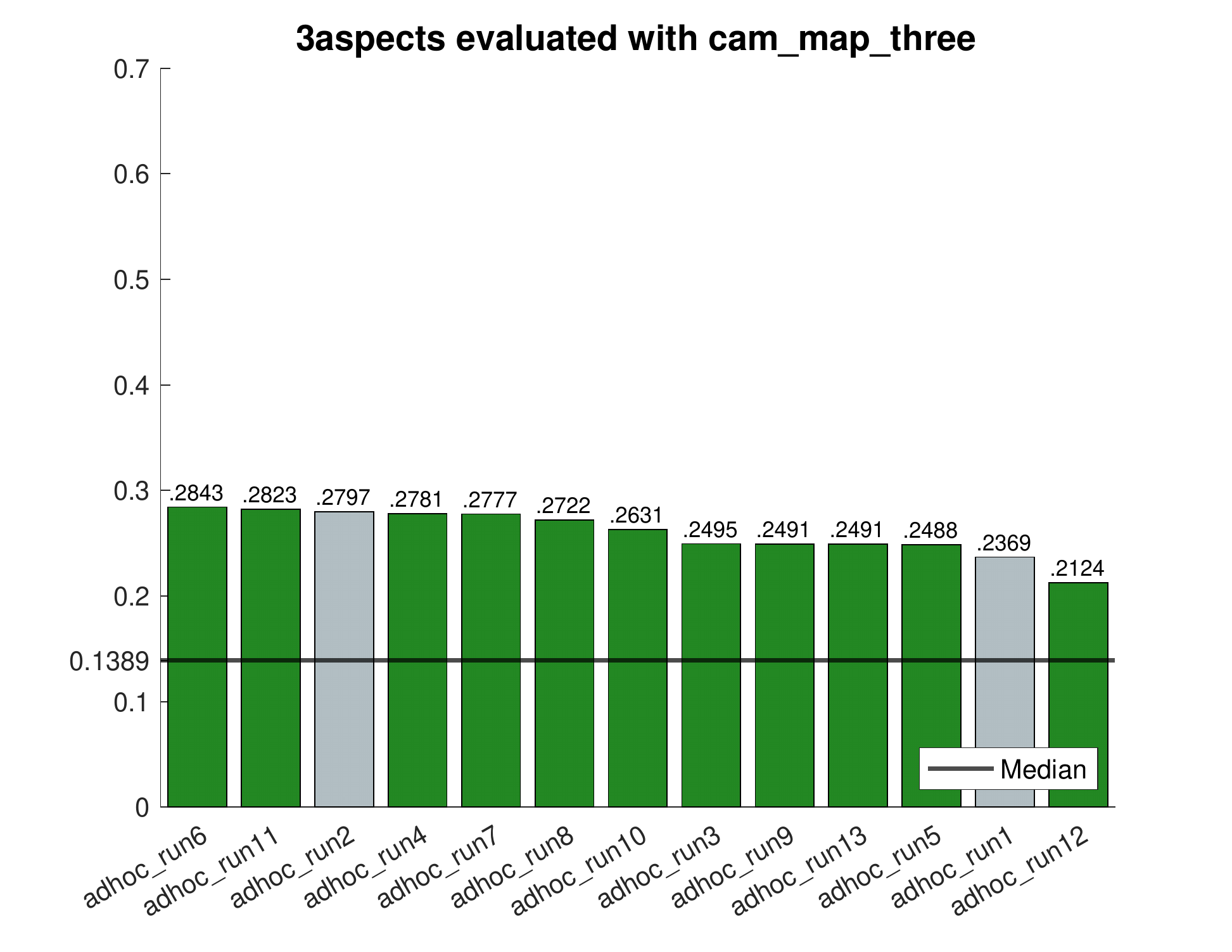}
  \caption{Useful, Correct and Credible, $\text{id}=2$}
  \label{fig:cam_useful_correct_credible}
\end{subfigure}
\caption{Ad Hoc Task: Our runs evaluated with \ac{CAM}.}
\label{fig:cam_evaluation}
\end{figure}

Figure~\ref{fig:cam_evaluation} shows ours runs evaluated with \ac{CAM} computed with \ac{AP}. Observe that the main difference between Figure~\ref{fig:binary_evaluation} and Figure~\ref{fig:cam_evaluation} is that in the former case document labels are aggregated across aspects and then mapped to a single binary label, while in the latter case each aspect is treated separately with \ac{AP} and then \ac{CAM} computes the average across aspects.

The best $3$ runs within all the sub-figures of Figure~\ref{fig:cam_evaluation} are: \texttt{adhoc\_run6}, \texttt{adhoc\_run11} and \texttt{adhoc\_run2}. \texttt{Adhoc\_run2} is one of our internal baseline, obtained with RM3 and without re-ranking. This might be due to the usefulness \ac{AP} score being much higher that the credibility and correctness \ac{AP} scores, thus leading \ac{CAM} to reward more runs which perform quite well in terms of usefulness.

A similar reason can explain why \texttt{adhoc\_run6} is the best run with respect to usefulness-credibility in Figure~\ref{fig:cam_useful_credible}, usefulness-credibility-correctness in Figure~\ref{fig:cam_useful_correct_credible}, and the second best run with respect to correctness and credibility in Figure~\ref{fig:cam_correct_credible}.

Finally, as observed for binary evaluation and the Total Recall Task, the RRF exploited by \texttt{adhoc\_run11} is an effective way to merge the runs obtained with relevance, credibility and reversed misinformation. Indeed, \texttt{adhoc\_run11} is the second best run for with respect to usefulness-credibility and usefulness-credibility-correctness in Figures~\ref{fig:cam_useful_correct_credible} and~\ref{fig:cam_useful_credible}, and the third best run with respect to correctness and credibility in Figure~\ref{fig:cam_correct_credible}.

\begin{figure}[b]
\centering
\begin{subfigure}{.49\textwidth}
  \centering
  \includegraphics[width=\linewidth]{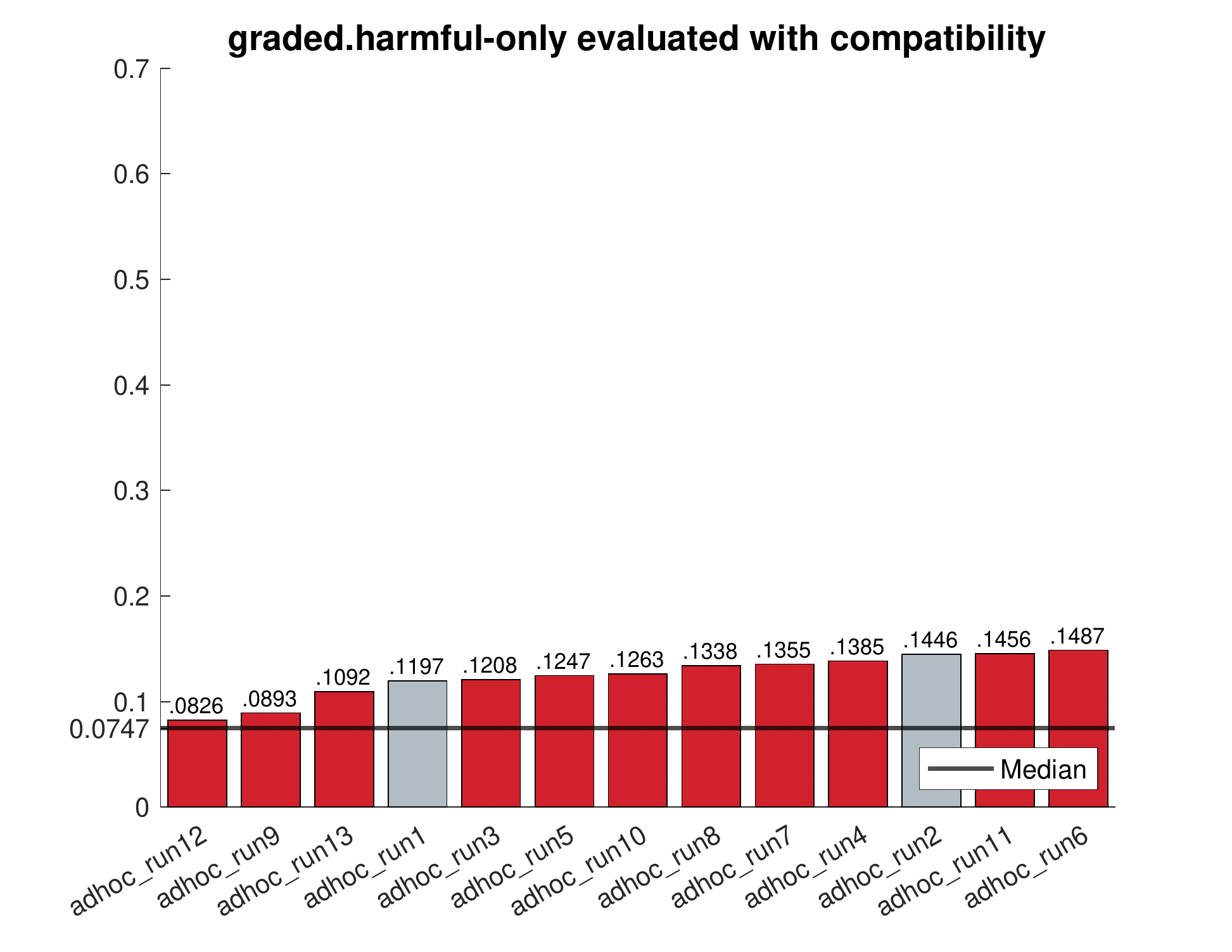}
  \caption{Harmful, $\text{id}=7$}
  \label{fig:harmful}
\end{subfigure}
\begin{subfigure}{.49\textwidth}
  \centering
  \includegraphics[width=\linewidth]{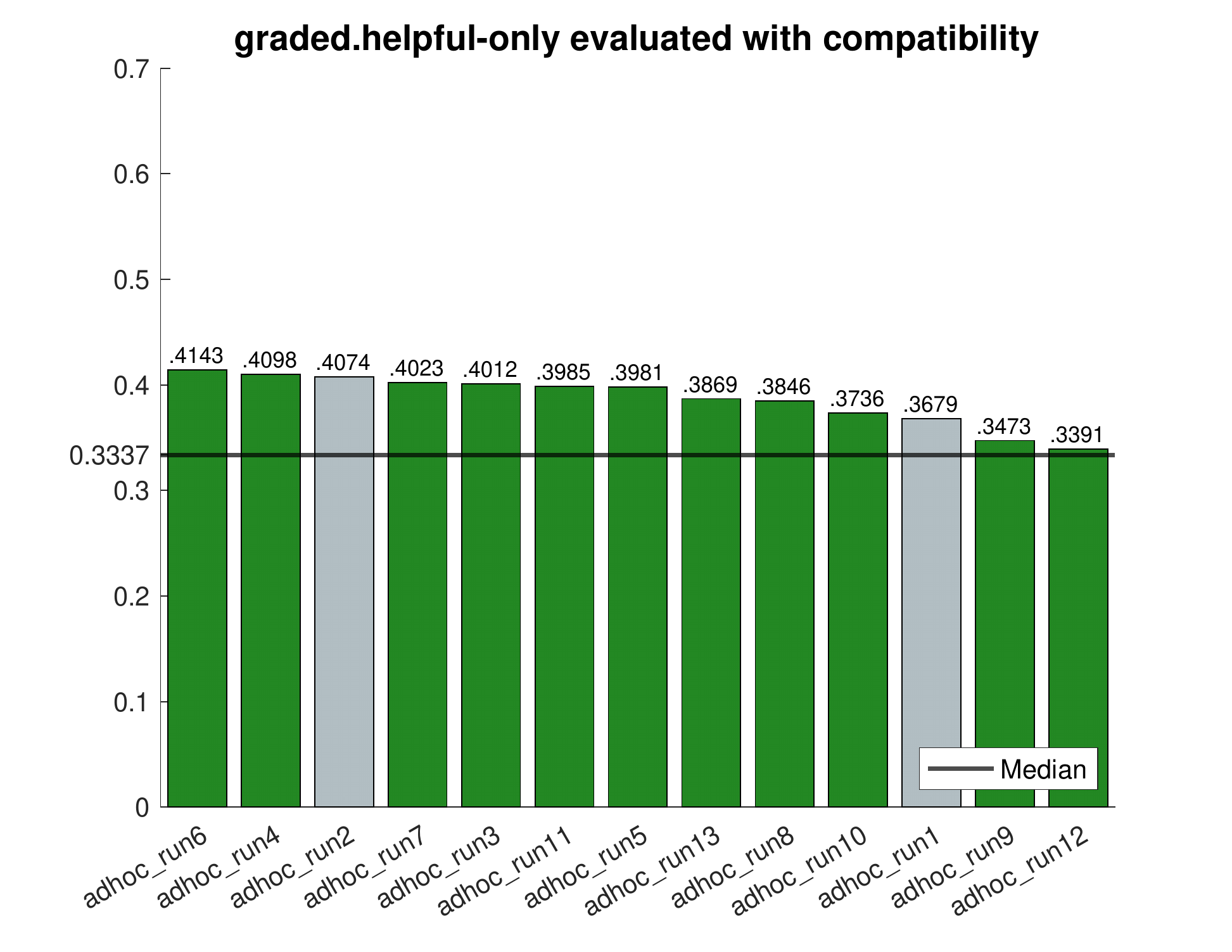}
  \caption{Helpful, $\text{id}=8$}
  \label{fig:helpful}
\end{subfigure}
\caption{Ad Hoc Task: Our runs evaluated with Compatibility.}
\label{fig:compatibility_evaluation}
\end{figure}

Figure~\ref{fig:compatibility_evaluation} shows harmfulness and helpfulness computed with compatibility. Recall that for harmfulness the lower the score the better the run, while for helpfulness the higher the score the better the run. As clearly illustrated in Figure~\ref{fig:harmful}, our runs are all worse than the task median for harmfulness, while they are all above the task median for helpfulness in Figure~\ref{fig:helpful}.

\begin{figure}[htb]
\centering
\includegraphics[width=0.5\linewidth]{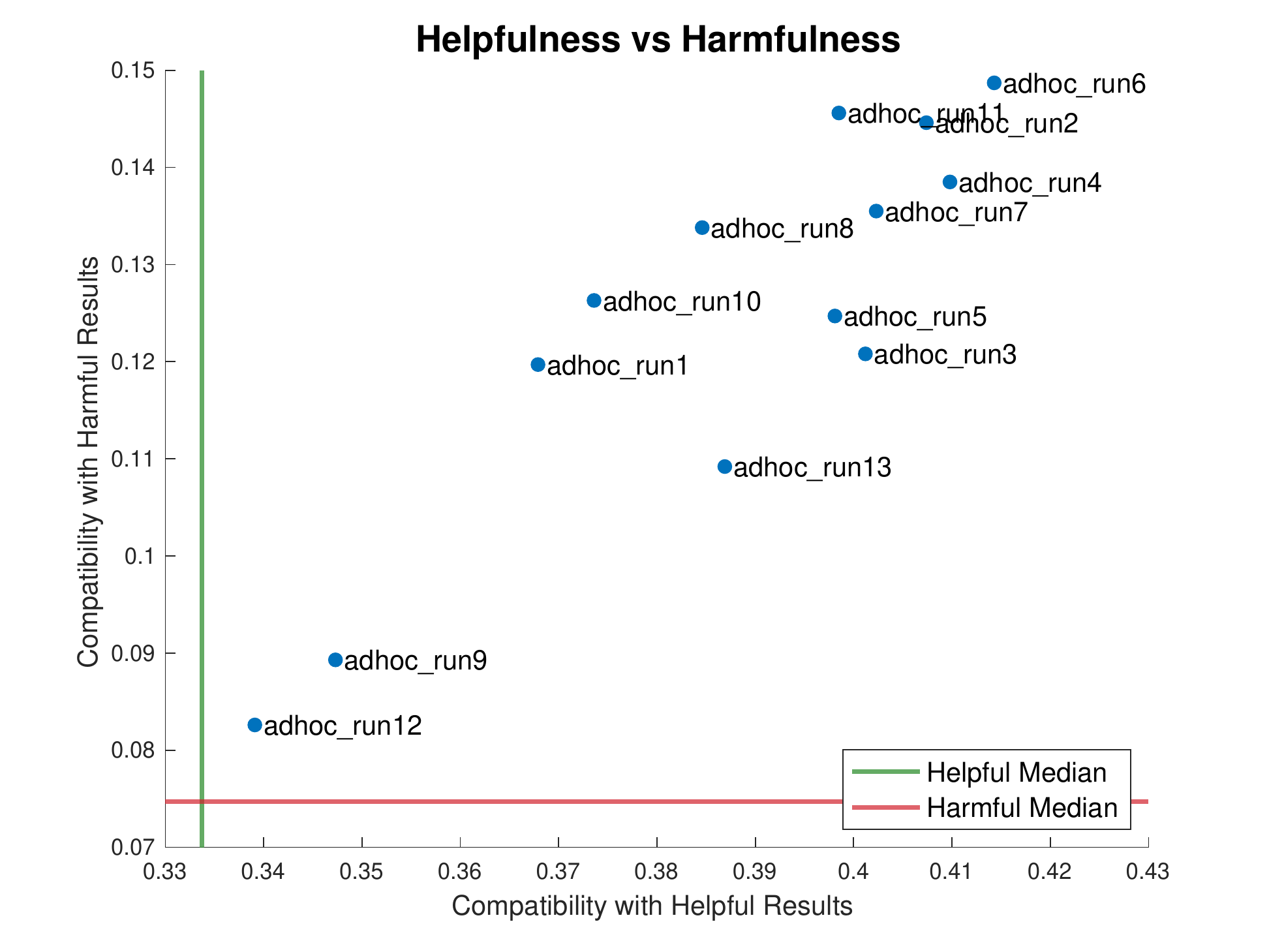}
\caption{Ad Hoc Task: Harmful scores against helpful scores. For harmful scores the lower the better.}
\label{fig:helpful_harmful}
\end{figure}

Figure~\ref{fig:helpful_harmful} plots harmfulness against helpfulness. The goal is to minimize harmfulness while maximizing helpfulness at the same time. Therefore, the closer a run is to the top left corner, the better the performance.

From Figure~\ref{fig:helpful_harmful}, \texttt{adhoc\_run13} is the closest run to the top left corner. \texttt{Adhoc\_run13} re-ranks the top $200$ documents just with the reversed misinformation score. As observed with the results of the Total Recall task, the misinformation score can be used to identify harmful documents and move them towards the end of the ranking. 

\texttt{adhoc\_run13} and \texttt{adhoc\_run10} are basically the same run, but \texttt{adhoc\_run13} re-ranks the top $200$ documents, while \texttt{adhoc\_run10} re-ranks the top $100$. This comparison suggests that if we consider higher cut-offs we might be able to reduce the harmfulness score even more.

In the bottom right corner of Figure~\ref{fig:helpful_harmful} there are \texttt{adhoc\_run12} and \texttt{adhoc\_run9}. These two runs are obtained by re-ranking the top $200$ and $100$ documents just with the credibility score. These are our best runs in terms of harmfulness. Again, by increasing the re-ranking cut-off, we decrease harmfulness.

A promising strategy to optimize harmfulness and helpfulness simultaneously might be to find a better way to aggregate the misinformation and credibility score. Indeed, from Figure~\ref{fig:helpful_harmful} all runs which aggregate the $3$ scores increase helpfulness, but are not able to decrease harmfulness.


\clearpage
\section{Related Work}
\label{sec:related}


\subsection{Automatic Classification of Credibility}


\citet{10.1007/978-3-642-36973-5_47} present a list of features used to train different machine learning models able to predict the credibility of Web pages. They divide those features in $2$ categories, features based on the content, as text readability, number of ads in the Web page, etc, and features based on social interactions, as number of shares on Facebook/Twitter, number of URLs, PageRank score, etc.

Similarly, \citet{KakolEtAl2017} present an automatic approach to classify the credibility of Web pages. The authors identify different features to classify credibility, as reading easiness, inclusion of links and references, language quality, the presence of author information, etc.  


\subsection{Automatic Fact Checking}

Automatic fact checking is typically either approached using propagation-based approaches, which determine the veracity of claims based on the spread of those claims in networks \cite{volkova2017separating,liu2019kernel,shu2019beyond,journals/corr/abs-1910-01363}, or, as in this paper, content-based ones \cite{thorne2018fever,augenstein2019multifc,conf/aaai/NieCB19}. 

Content-based approaches consist of multiple tasks: claim check-worthiness detection to identify relevant claims to fact-check \cite{atanasova2018overview,hansen2019neural,wright2020fact}; evidence document retrieval or reranking, if candidate documents are provided \cite{conf/emnlp/0001R18,allein2020timeaware}; stance detection to determine whether retrieved documents agree or disagree with the claim or neither of those \cite{conf/emnlp/AugensteinRVB16,conf/naacl/BalyMGMMN18,journals/corr/RiedelASR17,conf/naacl/FerreiraV16}; and finally, veracity prediction \cite{thorne2018fever,augenstein2019multifc,atanasova-etal-2020-generating}. 

For stance detection, which is used to rank misinformation in this work, commonly studied research challenges include the prediction of stances towards unseen targets \cite{conf/emnlp/AugensteinRVB16,journals/corr/abs-1902-02401}, involving multiple targets or evidence documents at the same time \cite{E17-2088,mohtarami-etal-2018-automatic,conf/ijcnn/WangWLCF20}, the unification of different stance label schemes \cite{conf/naacl/AugensteinRS18} and transfer learning \cite{journals/corr/abs-2001-01565}. State-of-the-art architectures for stance detection typically include large pre-trained Transformer models, as we also utilize here.

\section{Conclusions}
\label{sec:conclusions}

In this paper, we describe our participation in the TREC Health Misinformation Track 2020. We submitted $11$ runs to the Total Recall task and $13$ runs to the Ad Hoc task. Our approach consists of $3$ steps: (1) we create an initial run with BM25 or RM3; (2) we estimate credibility and misinformation scores for each document in the initial run; (3) we merged the scores across relevance, credibility and misinformation to re-rank documents. 

Experimental results show that all our runs are above the task median, except for harmfulness. However, when it comes to the estimation of credibility and misinformation, our approach seems promising. Therefore, to minimize harmfulness, we need to find a better way to identify harmful documents by aggregating misinformation and credibility scores. Indeed, as shown in the experimental results, approaches as the weighted average or RRF tend to optimize helpfulness at the expense of harmfulness.

\clearpage
\bibliographystyle{abbrvnat}

\bibliography{bibliography.bib}

\begin{thebibliography}{37}
\providecommand{\natexlab}[1]{#1}
\providecommand{\url}[1]{\texttt{#1}}
\expandafter\ifx\csname urlstyle\endcsname\relax
  \providecommand{\doi}[1]{doi: #1}\else
  \providecommand{\doi}{doi: \begingroup \urlstyle{rm}\Url}\fi

\bibitem[Allein et~al.(2020)Allein, Augenstein, and Moens]{allein2020timeaware}
L.~Allein, I.~Augenstein, and M.-F. Moens.
\newblock {Time-Aware Evidence Ranking for Fact-Checking}.
\newblock \emph{arXiv preprint arXiv:2009.06402}, 2020.

\bibitem[Atanasova et~al.(2018)Atanasova, Barron-Cedeno, Elsayed, Suwaileh,
  Zaghouani, Kyuchukov, Martino, and Nakov]{atanasova2018overview}
P.~Atanasova, A.~Barron-Cedeno, T.~Elsayed, R.~Suwaileh, W.~Zaghouani,
  S.~Kyuchukov, G.~D.~S. Martino, and P.~Nakov.
\newblock {Overview of the CLEF-2018 CheckThat! Lab on Automatic Identification
  and Verification of Political Claims. Task 1: Check-Worthiness}.
\newblock \emph{arXiv preprint arXiv:1808.05542}, 2018.

\bibitem[Atanasova et~al.(2020)Atanasova, Simonsen, Lioma, and
  Augenstein]{atanasova-etal-2020-generating}
P.~Atanasova, J.~G. Simonsen, C.~Lioma, and I.~Augenstein.
\newblock Generating fact checking explanations.
\newblock In \emph{Proceedings of the 58th Annual Meeting of the Association
  for Computational Linguistics}, pages 7352--7364, Online, July 2020.
  Association for Computational Linguistics.
\newblock \doi{10.18653/v1/2020.acl-main.656}.
\newblock URL \url{https://www.aclweb.org/anthology/2020.acl-main.656}.

\bibitem[Augenstein et~al.(2016{\natexlab{a}})Augenstein, Rockt{\"a}schel,
  Vlachos, and Bontcheva]{augenstein2016stance}
I.~Augenstein, T.~Rockt{\"a}schel, A.~Vlachos, and K.~Bontcheva.
\newblock {Stance Detection with Bidirectional Conditional Encoding}.
\newblock In \emph{Proceedings of the 2016 Conference on Empirical Methods in
  Natural Language Processing}, pages 876--885, 2016{\natexlab{a}}.

\bibitem[Augenstein et~al.(2016{\natexlab{b}})Augenstein, Rockt{\"a}schel,
  Vlachos, and Bontcheva]{conf/emnlp/AugensteinRVB16}
I.~Augenstein, T.~Rockt{\"a}schel, A.~Vlachos, and K.~Bontcheva.
\newblock {Stance Detection with Bidirectional Conditional Encoding}.
\newblock In J.~Su, X.~Carreras, and K.~Duh, editors, \emph{EMNLP}, pages
  876--885. The Association for Computational Linguistics, 2016{\natexlab{b}}.
\newblock ISBN 978-1-945626-25-8.
\newblock URL
  \url{http://dblp.uni-trier.de/db/conf/emnlp/emnlp2016.html#AugensteinRVB16}.

\bibitem[Augenstein et~al.(2018)Augenstein, Ruder, and
  Søgaard]{conf/naacl/AugensteinRS18}
I.~Augenstein, S.~Ruder, and A.~Søgaard.
\newblock Multi-task learning of pairwise sequence classification tasks over
  disparate label spaces.
\newblock In M.~A. Walker, H.~Ji, and A.~Stent, editors, \emph{NAACL-HLT},
  pages 1896--1906. Association for Computational Linguistics, 2018.
\newblock ISBN 978-1-948087-27-8.
\newblock URL
  \url{http://dblp.uni-trier.de/db/conf/naacl/naacl2018-1.html#AugensteinRS18}.

\bibitem[Augenstein et~al.(2019)Augenstein, Lioma, Wang, Lima, Hansen, Hansen,
  and Simonsen]{augenstein2019multifc}
I.~Augenstein, C.~Lioma, D.~Wang, L.~C. Lima, C.~Hansen, C.~Hansen, and J.~G.
  Simonsen.
\newblock {MultiFC: A Real-World Multi-Domain Dataset for Evidence-Based Fact
  Checking of Claims}.
\newblock In \emph{Proceedings of the 2019 Conference on Empirical Methods in
  Natural Language Processing and the 9th International Joint Conference on
  Natural Language Processing (EMNLP-IJCNLP)}, pages 4685--4697, Hong Kong,
  Nov. 2019. Association for Computational Linguistics.
\newblock \doi{10.18653/v1/D19-1475}.

\bibitem[Baly et~al.(2018)Baly, Mohtarami, Glass, Màrquez, Moschitti, and
  Nakov]{conf/naacl/BalyMGMMN18}
R.~Baly, M.~Mohtarami, J.~R. Glass, L.~Màrquez, A.~Moschitti, and P.~Nakov.
\newblock {Integrating Stance Detection and Fact Checking in a Unified Corpus}.
\newblock In M.~A. Walker, H.~Ji, and A.~Stent, editors, \emph{NAACL-HLT (2)},
  pages 21--27. Association for Computational Linguistics, 2018.
\newblock ISBN 978-1-948087-29-2.
\newblock URL
  \url{http://dblp.uni-trier.de/db/conf/naacl/naacl2018-2.html#BalyMGMMN18}.

\bibitem[Clarke et~al.(2020)Clarke, Maistro, Smucker, and
  Zuccon]{overview-misininfo-2020}
C.~Clarke, M.~Maistro, M.~Smucker, and G.~Zuccon.
\newblock {Overview of the TREC 2020 Health Misinformation Track (to appear)}.
\newblock In E.~M. Voorhees and A.~Ellis, editors, \emph{Proceedings of the
  Twenty-Nine Text REtrieval Conference, {TREC} 2020, Gaithersburg, Maryland,
  USA, November 16-19, 2020}, volume~- of \emph{{NIST} Special Publication}.
  National Institute of Standards and Technology {(NIST)}, 2020.

\bibitem[Cormack et~al.(2009)Cormack, Clarke, and Büttcher]{rrf}
G.~Cormack, C.~Clarke, and S.~Büttcher.
\newblock Reciprocal rank fusion outperforms condorcet and individual rank
  learning methods.
\newblock pages 758--759, 01 2009.
\newblock \doi{10.1145/1571941.1572114}.

\bibitem[Devlin et~al.(2019)Devlin, Chang, Lee, and Toutanova]{devlin2019bert}
J.~Devlin, M.-W. Chang, K.~Lee, and K.~Toutanova.
\newblock {BERT: Pre-training of Deep Bidirectional Transformers for Language
  Understanding}.
\newblock In \emph{NAACL-HLT}, 2019.

\bibitem[Ferreira and Vlachos(2016)]{conf/naacl/FerreiraV16}
W.~Ferreira and A.~Vlachos.
\newblock Emergent: a novel data-set for stance classification.
\newblock In K.~Knight, A.~Nenkova, and O.~Rambow, editors, \emph{HLT-NAACL},
  pages 1163--1168. The Association for Computational Linguistics, 2016.
\newblock ISBN 978-1-941643-91-4.
\newblock URL
  \url{http://dblp.uni-trier.de/db/conf/naacl/naacl2016.html#FerreiraV16}.

\bibitem[Hansen et~al.(2019)Hansen, Hansen, Alstrup, Grue~Simonsen, and
  Lioma]{hansen2019neural}
C.~Hansen, C.~Hansen, S.~Alstrup, J.~Grue~Simonsen, and C.~Lioma.
\newblock {Neural Check-Worthiness Ranking With Weak Supervision: Finding
  Sentences for Fact-Checking}.
\newblock In \emph{Companion Proceedings of the 2019 World Wide Web
  Conference}, pages 994--1000, 2019.

\bibitem[Hartmann et~al.(2019)Hartmann, Golovchenko, and
  Augenstein]{journals/corr/abs-1910-01363}
M.~Hartmann, Y.~Golovchenko, and I.~Augenstein.
\newblock Mapping (dis-)information flow about the mh17 plane crash.
\newblock \emph{CoRR}, abs/1910.01363, 2019.
\newblock URL
  \url{http://dblp.uni-trier.de/db/journals/corr/corr1910.html#abs-1910-01363}.

\bibitem[Kakol et~al.(2017)Kakol, Nielek, and Wierzbicki]{KakolEtAl2017}
M.~Kakol, R.~Nielek, and A.~Wierzbicki.
\newblock {Understanding and Predicting Web Content Credibility Using the
  Content Credibility Corpus}.
\newblock \emph{Information Processing \& Management}, 53\penalty0
  (5):\penalty0 1043--1061, 2017.

\bibitem[Lavrenko and Croft(2001)]{LavrenkoAndCroft}
V.~Lavrenko and W.~B. Croft.
\newblock Relevance based language models.
\newblock In \emph{Proceedings of the 24th Annual International ACM SIGIR
  Conference on Research and Development in Information Retrieval}, SIGIR '01,
  pages 120--127, New York, NY, USA, 2001. Association for Computing Machinery.
\newblock ISBN 1581133316.

\bibitem[Lioma et~al.(2016)Lioma, Larsen, Lu, and
  Huang]{10.1145/3006299.3006315}
C.~Lioma, B.~Larsen, W.~Lu, and Y.~Huang.
\newblock A study of factuality, objectivity and relevance: Three desiderata in
  large-scale information retrieval?
\newblock In \emph{Proceedings of the 3rd IEEE/ACM International Conference on
  Big Data Computing, Applications and Technologies}, BDCAT '16, page
  107–117, New York, NY, USA, 2016. Association for Computing Machinery.
\newblock ISBN 9781450346177.
\newblock \doi{10.1145/3006299.3006315}.
\newblock URL \url{https://doi.org/10.1145/3006299.3006315}.

\bibitem[Lioma et~al.(2017)Lioma, Simonsen, and
  Larsen]{10.1145/3121050.3121072}
C.~Lioma, J.~G. Simonsen, and B.~Larsen.
\newblock Evaluation measures for relevance and credibility in ranked lists.
\newblock In \emph{Proceedings of the ACM SIGIR International Conference on
  Theory of Information Retrieval}, ICTIR '17, page 91–98, New York, NY, USA,
  2017. Association for Computing Machinery.
\newblock ISBN 9781450344906.
\newblock \doi{10.1145/3121050.3121072}.
\newblock URL \url{https://doi.org/10.1145/3121050.3121072}.

\bibitem[Liu et~al.(2019)Liu, Ott, Goyal, Du, Joshi, Chen, Levy, Lewis,
  Zettlemoyer, and Stoyanov]{liu2019roberta}
Y.~Liu, M.~Ott, N.~Goyal, J.~Du, M.~Joshi, D.~Chen, O.~Levy, M.~Lewis,
  L.~Zettlemoyer, and V.~Stoyanov.
\newblock {Roberta: A robustly optimized bert pretraining approach}.
\newblock \emph{arXiv preprint arXiv:1907.11692}, 2019.

\bibitem[Liu et~al.(2020)Liu, Xiong, Sun, and Liu]{liu2019kernel}
Z.~Liu, C.~Xiong, M.~Sun, and Z.~Liu.
\newblock Fine-grained fact verification with kernel graph attention network.
\newblock In \emph{Proceedings of the 58th Annual Meeting of the Association
  for Computational Linguistics}, pages 7342--7351, Online, July 2020.
  Association for Computational Linguistics.
\newblock \doi{10.18653/v1/2020.acl-main.655}.

\bibitem[Mohammad et~al.(2016)Mohammad, Kiritchenko, Sobhani, Zhu, and
  Cherry]{mohammad2016semeval}
S.~Mohammad, S.~Kiritchenko, P.~Sobhani, X.~Zhu, and C.~Cherry.
\newblock {SemEval-2016 Task 6: Detecting Stance in Tweets}.
\newblock In \emph{Proceedings of the 10th International Workshop on Semantic
  Evaluation (SemEval-2016)}, pages 31--41, 2016.

\bibitem[Mohtarami et~al.(2018)Mohtarami, Baly, Glass, Nakov, M{\`a}rquez, and
  Moschitti]{mohtarami-etal-2018-automatic}
M.~Mohtarami, R.~Baly, J.~Glass, P.~Nakov, L.~M{\`a}rquez, and A.~Moschitti.
\newblock {Automatic Stance Detection Using End-to-End Memory Networks}.
\newblock In \emph{Proceedings of the 2018 Conference of the North {A}merican
  Chapter of the Association for Computational Linguistics: Human Language
  Technologies, Volume 1 (Long Papers)}, pages 767--776, New Orleans,
  Louisiana, June 2018. Association for Computational Linguistics.
\newblock \doi{10.18653/v1/N18-1070}.
\newblock URL \url{https://www.aclweb.org/anthology/N18-1070}.

\bibitem[Nie et~al.(2019)Nie, Chen, and Bansal]{conf/aaai/NieCB19}
Y.~Nie, H.~Chen, and M.~Bansal.
\newblock {Combining Fact Extraction and Verification with Neural Semantic
  Matching Networks}.
\newblock In \emph{AAAI}, pages 6859--6866. AAAI Press, 2019.
\newblock ISBN 978-1-57735-809-1.
\newblock URL
  \url{http://dblp.uni-trier.de/db/conf/aaai/aaai2019.html#NieCB19}.

\bibitem[Olteanu et~al.(2013)Olteanu, Peshterliev, Liu, and
  Aberer]{10.1007/978-3-642-36973-5_47}
A.~Olteanu, S.~Peshterliev, X.~Liu, and K.~Aberer.
\newblock Web credibility: Features exploration and credibility prediction.
\newblock In P.~Serdyukov, P.~Braslavski, S.~O. Kuznetsov, J.~Kamps,
  S.~R{\"u}ger, E.~Agichtein, I.~Segalovich, and E.~Yilmaz, editors,
  \emph{Advances in Information Retrieval}, pages 557--568, Berlin, Heidelberg,
  2013. Springer Berlin Heidelberg.
\newblock ISBN 978-3-642-36973-5.

\bibitem[Riedel et~al.(2017)Riedel, Augenstein, Spithourakis, and
  Riedel]{journals/corr/RiedelASR17}
B.~Riedel, I.~Augenstein, G.~P. Spithourakis, and S.~Riedel.
\newblock {A simple but tough-to-beat baseline for the Fake News Challenge
  stance detection task}.
\newblock \emph{CoRR}, abs/1707.03264, 2017.
\newblock URL
  \url{http://dblp.uni-trier.de/db/journals/corr/corr1707.html#RiedelASR17}.

\bibitem[Schiller et~al.(2020)Schiller, Daxenberger, and
  Gurevych]{journals/corr/abs-2001-01565}
B.~Schiller, J.~Daxenberger, and I.~Gurevych.
\newblock {Stance Detection Benchmark: How Robust Is Your Stance Detection?}
\newblock \emph{CoRR}, abs/2001.01565, 2020.
\newblock URL
  \url{http://dblp.uni-trier.de/db/journals/corr/corr2001.html#abs-2001-01565}.

\bibitem[Schwarz and Morris(2011)]{schwarz2011augmenting}
J.~Schwarz and M.~R. Morris.
\newblock Augmenting web pages and search results to support credibility
  assessment.
\newblock In \emph{ACM Conference on Computer-Human Interaction}, May 2011.
\newblock URL
  \url{https://www.microsoft.com/en-us/research/publication/augmenting-web-pages-and-search-results-to-support-credibility-assessment/}.

\bibitem[Shu et~al.(2019)Shu, Wang, and Liu]{shu2019beyond}
K.~Shu, S.~Wang, and H.~Liu.
\newblock Beyond news contents: The role of social context for fake news
  detection.
\newblock In \emph{Proceedings of the 12th ACM International Conference on Web
  Search and Data Mining}, pages 312--320, New York, 2019. Association for
  Computing Machinery.
\newblock \doi{10.1145/3289600.3290994}.

\bibitem[Slovikovskaya and Attardi(2020)]{slovikovskaya2020transfer}
V.~Slovikovskaya and G.~Attardi.
\newblock {Transfer Learning from Transformers to Fake News Challenge Stance
  Detection (FNC-1) Task}.
\newblock In \emph{Proceedings of The 12th Language Resources and Evaluation
  Conference}, pages 1211--1218, 2020.

\bibitem[Sobhani et~al.(2017)Sobhani, Inkpen, and Zhu]{E17-2088}
P.~Sobhani, D.~Inkpen, and X.~Zhu.
\newblock {A Dataset for Multi-Target Stance Detection}.
\newblock In \emph{Proceedings of the 15th Conference of the European Chapter
  of the Association for Computational Linguistics: Volume 2, Short Papers},
  pages 551--557. Association for Computational Linguistics, 2017.
\newblock URL \url{http://aclweb.org/anthology/E17-2088}.

\bibitem[Thorne et~al.(2018)Thorne, Vlachos, Christodoulopoulos, and
  Mittal]{thorne2018fever}
J.~Thorne, A.~Vlachos, C.~Christodoulopoulos, and A.~Mittal.
\newblock {FEVER: a Large-scale Dataset for Fact Extraction and VERification}.
\newblock In \emph{Proceedings of the 2018 Conference of the North American
  Chapter of the Association for Computational Linguistics: Human Language
  Technologies, Volume 1 (Long Papers)}, pages 809--819, 2018.

\bibitem[Volkova et~al.(2017)Volkova, Shaffer, Jang, and
  Hodas]{volkova2017separating}
S.~Volkova, K.~Shaffer, J.~Y. Jang, and N.~Hodas.
\newblock Separating facts from fiction: Linguistic models to classify
  suspicious and trusted news posts on twitter.
\newblock In \emph{Proceedings of the 55th Annual Meeting of the Association
  for Computational Linguistics (Volume 2: Short Papers)}, pages 647--653,
  Vancouver, 2017. Association for Computational Linguistics.
\newblock \doi{10.18653/v1/P17-2102}.

\bibitem[Wang et~al.(2020)Wang, Wang, Lv, Cao, and Fu]{conf/ijcnn/WangWLCF20}
Z.~Wang, Q.~Wang, C.~Lv, X.~Cao, and G.~Fu.
\newblock {Unseen Target Stance Detection with Adversarial Domain
  Generalization}.
\newblock In \emph{IJCNN}, pages 1--8. IEEE, 2020.
\newblock ISBN 978-1-7281-6926-2.
\newblock URL
  \url{http://dblp.uni-trier.de/db/conf/ijcnn/ijcnn2020.html#WangWLCF20}.

\bibitem[Wolf et~al.(2019)Wolf, Debut, Sanh, Chaumond, Delangue, Moi, Cistac,
  Rault, Louf, Funtowicz, Davison, Shleifer, von Platen, Ma, Jernite, Plu, Xu,
  Scao, Gugger, Drame, Lhoest, and Rush]{Wolf2019HuggingFacesTS}
T.~Wolf, L.~Debut, V.~Sanh, J.~Chaumond, C.~Delangue, A.~Moi, P.~Cistac,
  T.~Rault, R.~Louf, M.~Funtowicz, J.~Davison, S.~Shleifer, P.~von Platen,
  C.~Ma, Y.~Jernite, J.~Plu, C.~Xu, T.~L. Scao, S.~Gugger, M.~Drame, Q.~Lhoest,
  and A.~M. Rush.
\newblock {HuggingFace's Transformers: State-of-the-art Natural Language
  Processing}.
\newblock \emph{ArXiv}, abs/1910.03771, 2019.

\bibitem[Wright and Augenstein(2020)]{wright2020fact}
D.~Wright and I.~Augenstein.
\newblock {Claim Check-Worthiness Detection as Positive Unlabelled Learning}.
\newblock In \emph{Findings of EMNLP}. Association for Computational
  Linguistics, 2020.

\bibitem[Xu et~al.(2019)Xu, Mohtarami, and Glass]{journals/corr/abs-1902-02401}
B.~Xu, M.~Mohtarami, and J.~R. Glass.
\newblock {Adversarial Domain Adaptation for Stance Detection}.
\newblock \emph{CoRR}, abs/1902.02401, 2019.
\newblock URL
  \url{http://dblp.uni-trier.de/db/journals/corr/corr1902.html#abs-1902-02401}.

\bibitem[Yin and Roth(2018)]{conf/emnlp/0001R18}
W.~Yin and D.~Roth.
\newblock {TwoWingOS: A Two-Wing Optimization Strategy for Evidential Claim
  Verification}.
\newblock In E.~Riloff, D.~Chiang, J.~Hockenmaier, and J.~Tsujii, editors,
  \emph{EMNLP}, pages 105--114. {Association for Computational Linguistics},
  2018.
\newblock ISBN 978-1-948087-84-1.
\newblock URL
  \url{http://dblp.uni-trier.de/db/conf/emnlp/emnlp2018.html#0001R18}.

\end{thebibliography}

\end{document}